\documentclass[11pt]{article}

\usepackage{fourier}
\usepackage[utf8]{inputenc}
\usepackage{booktabs}
\usepackage{multirow}
\usepackage{graphicx}
\usepackage{float}
\usepackage{apacite}
\usepackage[margin=1.2in]{geometry}
\usepackage{amsmath}
\usepackage{comment}
\usepackage{bm,xcolor,bbm,authblk}
\usepackage[symbol]{footmisc}
\usepackage{relsize}

\makeatletter
\def\@fnsymbol#1{\ensuremath{\ifcase#1\or \sharp\or \partial\or \natural\or \flat \else\@ctrerr\fi}}
\makeatother

\newcommand*{\grey}{\textcolor{gray}}
\newcommand{\norm}[1]{\lvert #1 \rvert}

\title{Using leave-one-out cross-validation (LOO) in a multilevel regression and poststratification (MRP) workflow: A cautionary tale}

\author{Swen Kuh\footnote{Department of Econometrics and Business Statistics, Monash University} \hspace{1em} Lauren Kennedy$^\sharp$ \hspace{0.8em} Qixuan Chen\footnote{Department of Biostatistics, Columbia University} \hspace{1em} Andrew Gelman\footnote{Department of Statistics and Political Science, Columbia University} }
\date{29 Aug 2022}

\begin{document}

\maketitle

\begin{abstract}
In recent decades, multilevel regression and poststratification (MRP) has surged in popularity for population inference. However, the validity of the estimates can depend on details of the model, and there is currently little research on validation. We explore how leave-one-out cross-validation (LOO) can be used to compare Bayesian models for MRP.  We investigate two approximate calculations of LOO, the Pareto smoothed importance sampling (PSIS-LOO) and a survey-weighted alternative (WTD-PSIS-LOO). Using two simulation designs, we examine how accurately these two criteria recover the correct ordering of model goodness at predicting population and small area level estimands. Focusing first on variable selection, we find that neither PSIS-LOO nor WTD-PSIS-LOO correctly recovers the models’ order for an MRP population estimand (although both criteria correctly identify the best and worst model). When considering small-area estimation, the best model differs for different small areas, highlighting the complexity of MRP validation. When considering different priors, the models' order seems slightly better at smaller area levels. These findings suggest that while not terrible, PSIS-LOO-based ranking techniques may not be suitable to evaluate MRP as a method. We suggest this is due to the aggregation stage of MRP, where individual-level prediction errors average out. These results show that in practice, PSIS-LOO-based model validation tools need to be used with caution and might not convey the full story when validating MRP as a method.
\end{abstract}

\section{Introduction}
Multilevel regression and poststratification (MRP) is a method reliant on models, so it is essential to have a method to validate our models. MRP has experienced a surge in popularity since its first proposal \cite{gelman1997}. This method is formed through two stages: the first stage requires a regularized model, which is used to directly model the outcome that we would like to estimate in the population by postratification variables. The second uses this model to predict this outcome into the population and then aggregate over these predictions to make population and small area estimates. As its popularity has grown, so too have the successful application areas \cite{park2004, lax2009gay, wang2015, downes2018}. 
However, MRP as a method is beholden to the goodness of the model. We see this in the initial criticism by \citeA{valliant2020} and subsequent reanalysis by \citeA{si2017} that the goodness of the estimand is highly dependent on the goodness of the model. However, just how do we validate the goodness of the model, and which models should be included for comparison?

In this manuscript, we focus on two types of candidate models for comparisons --- different variables and different priors. With the first model comparison, we consider which variables should be included in the model. We categorize variables into different types based on their impact on the estimand and work under the framework that a model designed for an MRP method should be concise (only include the most necessary variables). The second model comparison focuses on holding the variables constant but considering different forms of regularization instead.

To compare the goodness of the models, we focus on two methods. Firstly, we use an approximation to LOO, the  Pareto smoothed importance sampling (PSIS-LOO) \cite{vehtari2017} as implemented in the \texttt{loo} package in R, and secondly, a weighted alternative of this approximation (WTD-PSIS-LOO). These methods were chosen for their practical speed (as they can be used without refitting the models), and their overall popularity (suggesting a high likeliness that they are already being used to validate MRP models). However, we demonstrate that there is an inherent challenge to this --- while PSIS-LOO recovers the order of models well for \textit{individual}-level estimates, at a wider \textit{population} or small area estimates, this ordering is not maintained. Although we find some reassurance that the best model is consistently identified, we suggest that challenges around ordering suggest researchers and practitioners should proceed with caution when using LOO for MRP.
\vspace{1em}
\subsection{What is MRP?}
MRP \cite{gelman1997,park2004, park2006} was first proposed in a political science context \cite{lax2009gay, warshaw2012}. Since then, it has continued to be widely applied in political science \cite{lax2009should, buttice2013, ghitza2013}, and expanded to other fields such as health sciences \cite{zhang2015,downes2018}, survey research \cite{si2017} and most recently in inferring the prevalence and transmission of the COVID-19 pandemic \cite{covello2021}. This popularity brings two salient obstacles for practitioners who seek to use MRP credibly. Firstly, MRP does not have a single goal but rather two --- it is used to make population-level and small-area estimations. Secondly, the diverse range of applications suggests that the models will differ based on context. 

For this manuscript, we distinguish between the method of MRP and the model used in MRP.  The method of MRP has two stages. First, a multilevel model (the model used in MRP) is fitted to a sample with an outcome of interest and predictive categorical variables. Then, in the second stage,  we poststratify the posterior quantities from the first stage to the corresponding population, resulting in a posterior distribution for $\theta$. Usually, the model is a multilevel regression (hence the namesake multilevel regression and poststratification), but other predictive models can also be used (e.g., \citeA{gao2021treatment, ornstein2020, bisbee2019, Liu2020, liu2022}).

We follow the following steps using notation similar to \citeA{gelman2006} (as cited in \citeA{gao2021}): \\ 

1) \textbf{Multilevel regression:} 

 \begin{align}
     Pr(y_i = 1) &= \text{logit}^{-1}\left( \sum_{k = 1}^K \alpha^{(k)}_{\ell[i]} \right) \quad \text{for } i = 1, \ldots, n \label{eq:mlr1} \\
     \alpha^{(k)}_{\ell[i]} | \sigma^{(k)} &\sim \mbox{normal}(0, \sigma^{(k)}) \hspace{2.5em} \text{for } k = 1, \ldots, K; \ \ell^{(k)} = 1, \ldots, L^{(k)} \label{eq:mlr2}\\ 
     \sigma^{(k)} &\sim \mbox{normal}^+(0,1) \hspace{3.3em} \text{for } k = 1, \ldots, K \label{eq:mlr3}
 \end{align}
 \noindent where $y_i$ is the outcome for each individual $i$, $k$ is the $k$th predictor variable for the total $K$ number of variables, $\ell^{(k)}$ is the variable level and $L^{(k)}$ is the total number of levels for the $k$th variable. We use the symbol $\ell$ for small area level, not to be confused with $j$ as the corresponding number of cells on the poststratification table.
 
2) \textbf{Poststratification:} The predictions are made for each combination of the levels in the variables (\citeA{gelman2006}). The variables often represent geographic and demographic descriptors, but not necessarily (e.g., age groups, income level and residential postcode, etc.). This matrix of all the combinations of levels is called a poststratification table. These predictions are then aggregated to obtain a population level estimand by using known population sizes $N_j$ (usually from census or a large-scale survey at the national level) for every $j^{th}$ cell in the postratification matrix

\begin{equation}\label{eq:poststrat}
    \theta := \frac{\Sigma_{j \in J} N_j \theta_j}{\Sigma_{j \in J} N_j}.
\end{equation}

To extend this formula to small area estimation, we define $S$ as a particular category or level of the variable defined based on the poststratification table. In limit, $S=J$, the full population. For population-level estimates, the sum of all cell sizes amounts to the population size $\Sigma_{j \in S} N_j = N$, while for small area estimation, the sum of all cell sizes would be the population size for the particular category of a variable of interest $\Sigma_{j \in S} N_j = N_S$. The goal quantity $\theta_S$ would correspond to the proportion of individuals ``in the sub-population that would respond yes to the survey question of interest'' \cite{gao2021}. 

\begin{equation}\label{eq:poststrat_sae}
    \theta_S := \frac{\Sigma_{j \in S} N_j \theta_j}{\Sigma_{j \in S} N_j}.
\end{equation}

Although the two steps of the MRP method are often discussed separately, they are not independent. To poststratify, we require all the adjustment variables in the model to be known in the population (or at least known counts of different combinations). If they are not, work is required to impute these variables in the population (\citeA{kastellec2015polarizing, si2017, kuriwaki2021geography}). This is not always possible, and where it requires substantial work to find data and build an imputation technique. If we could correctly identify the variables that are needed in the model for the MRP method, this would mean this work is only done where it is absolutely needed. 

One of the strengths of the MRP method is the use of regularizing models. The regularizing model is particularly useful as it enables us to create estimates of groups with relatively small samples. Regularization is powerful in that it allows greater estimate stability and precision, but it also can create bias in estimands. Regularization is a feature of the model, but how can we identify the correct amount (tightness of prior) and type (e.g., a varying effect prior instead of an autoregressive prior)? If the regularization is too strong or inappropriate for the data and the sample size or size of a particular small area is small, the model will not correct this. 

\subsection{Challenges with validating MRP models}
Assessing the model fit and checking the model are crucial to statistical analysis \cite{gelman2013}. In previous sections, we have differentiated between the MRP method (model, predict and average) and the MRP model. Here we focus specifically on how to validate a model, and in particular how to validate a model in terms of predictive ability. We are not the first to think of validating the MRP model in terms of predictive ability. \citeA{wang2015difficulty} discuss the use of $k$-fold cross-validation for multilevel models in a survey context. They note the difficulty of identifying where a partially pooled model is best suited for the data, as well as noting the difficulty of identifying which folds to take in cross-validation. Our work builds on this existing work by focusing on validating MRP the process (not just the model), as well as the use of leave-one-out cross-validation (see Section \ref{ssec:whyloo}). 

Although \citeA{wang2015difficulty} do not explicitly discuss MRP, their focus on electoral survey predictions and simultaneous publication of work using MRP to make electoral predictions using highly unrepresentative data (e.g., the Xbox study \cite{wang2015}) suggests that validating the MRP model was perhaps their underlying goal. In addition to the challenges of validating a multilevel model, we also argue there are MRP-specific challenges. We discuss three main challenges with validating MRP models.

When we began this work we believed that the biggest challenge would be differences between sample and population data. Naturally, if we are interested in applying an MRP method, there is an implicit assumption that the sample is not a random sample, and that the probability of being included in the sample is potentially related with the quantity we would like to estimate. In this work, we combine the computationally efficient PSIS-LOO method proposed by \citeA{vehtari2017} with the weighted cross-validation proposed by \citeA{lumley2015} to address this issue. 

However, later in the manuscript, we will argue that there is another challenge for the validation of the MRP method. That is, it is not individual predictions that need to be good, but rather the aggregations of these individual estimates. We argue that this means the cost (or utility) function (expected log predictive density (elpd)) employed by both \citeA{vehtari2017} and \citeA{wang2015difficulty} is better used to evaluate the goodness of individual predictions than a method that requires aggregation like MRP. 

The aggregation challenge is also seen in variable selection in causal inference. This should not be surprising, \citeA{mercer2017theory} argue that there are several similarities between causal inference and survey adjustment. Of particular note is the distinction between bias variables (variables that if unadjusted for result in bias in estimates of the average treatment effect) and precision variables (variables that if unadjusted for result in less precise estimates of the average treatment effect). Similar variables distinctions are demonstrated in survey adjustment by \citeA{little2003} and used in our first simulation study. While in causal inference, solutions like double-LASSO \cite{urminsky2016using} effectively address this, this method is less suited to an MRP context due to the lack of information known about the population. 

The final challenge is that the best model for MRP depends on what the goals of the method are (population versus small area estimation). We demonstrate differences in the best model between different small areas. Although we do not claim a comprehensive solution to all three challenges, we demonstrate the efficacy of WTD-PSIS-LOO to solve the first challenge (the lack of random sampling), and argue that the underlying cause of the second (aggregation) is the use of elpd as a cost function (and discuss alternatives in the discussion) and demonstrate that clarity of the MRP goal is essential to correctly identify the best model. 
 
\subsection{Why LOO for MRP?}\label{ssec:whyloo}

There are two primary drivers for focusing this work on leave-one-out (LOO) cross-validation instead of $k$-fold cross-validation. The first is entirely pragmatic. The availability of a fast and effective approximate estimate for leave-one-out cross-validation (PSIS-LOO) proposed by \citeA{vehtari2017} and made accessible through the \texttt{loo} package (\citeA{vehtari2021package}). This makes PSIS-LOO practically appealing, especially given the lengthy computation time for complex MRP models. The second is the ability to easily incorporate survey weights into this approximation to adjust for sample representation (WTD-PSIS-LOO). 

We investigate LOO in our MRP work as the first stage of an MRP method is usually a Bayesian multilevel model to predict at the population level. LOO is often used to compare and validate predictive Bayesian models, due to its attractiveness and simplicity. On top of that, \citeA{vehtari2017} propose PSIS-LOO as a faster approximation to LOO as fitting complex models such as a Bayesian multilevel model is computationally expensive. LOO and PSIS-LOO are essentially estimates to the elpd, as a measure of predictive accuracy of models. \citeA{vehtari2017} define the elpd for a new dataset as

$$\text{elpd} = \sum_{i = 1}^n \int p_t(\tilde{y}_i) \text{ log } p(\tilde{y}_i | y) d\tilde{y}_i $$

\noindent where $p_t(\tilde{y}_i)$ is the distribution of the true data-generating process for $\tilde{y}_i$. While PSIS-LOO defined by \citeA{vehtari2017} takes the form of 

\begin{align}
    \widehat{\text{elpd}}_{\text{PSIS-LOO}} &= \mathlarger{\mathlarger{\sum}}_{i = 1}^n \ \text{log  } \ \underbrace{\frac{\sum_{s = 1}^S v_i^s p(y_i | \bm{\delta}^S)}{\sum_{s = 1}^S v_i^s}}_\text{$f^*_{\bm{\delta}}(y_i | \bm{x}_i)$} \label{eq:psis1} 
\end{align}
\noindent where $v_i^s$ is the replaced notation for smoothed weights using the Pareto distribution for a long-tailed weight distribution, $\bm{\delta}^S$ is the replaced notation for sampled posterior parameters $\bm{\delta}$ for samples $s = 1, \ldots, S$. We replace some of the notations from \citeA{vehtari2017} to be consistent with the work in this manuscript. For more details, see \citeA{vehtari2017}.

The model with the largest     $\widehat{\text{elpd}}_{\text{PSIS-LOO}}$ value is usually chosen to be the selected model, as a high value of posterior predictive density value indicates the new observation is well-accounted by the model. The main goal of the MRP model is to make predictions at the population level, or sometimes at the small area level. However, samples used in MRP are often unrepresentative of the population, so we propose a weighted alternative, as suggested in \citeA{lumley2015}:

\begin{align}
    \widehat{\text{elpd}}_{\text{WTD-LOO}} = \frac{1}{n} \sum_{i = 1}^n w_i \text{ log } f_{\bm{\delta}}(y_i | \bm{x}_i) \label{eq:wtdloo}
\end{align} 
where $w_i$ is the survey weight for each individual $i$ and $f_{\bm{\delta}}(y_i | \bm{x}_i)$ is the conditional density of $y$ given $\bm{x}$ with the parameters $\bm{\delta}$. 

Thus, our proposed weighted alternative WTD-PSIS-LOO takes the final form of

\begin{align}
    \widehat{\text{elpd}}_{\text{WTD-PSIS-LOO}} = \frac{1}{n} \mathlarger{\mathlarger{\sum}}_{i = 1}^n w_i \text{ log } \frac{\sum_{s = 1}^S v_i^s p(y_i | \bm{\delta}^S)}{\sum_{s = 1}^S v_i^s} \label{eq:wtdpsis}
\end{align} 

\noindent where we replace $f_{\bm{\delta}}(y_i | \bm{x}_i)$ in equation (\ref{eq:wtdloo}) with the Pareto-weighted density ($f^*_{\bm{\delta}}(y_i | \bm{x}_i)$) in equation (\ref{eq:psis1}). We calculate both WTD-PSIS-LOO and the unadjusted PSIS-LOO and compare their efficacy as a measure of predictive accuracy in our simulation studies. 

\subsection{How do we judge the MRP method?}

So what is a good MRP model? As MRP methods consist of two stages, it is not as straightforward as simply finding a good model independent of the poststratification step. Instead, we focus on the estimand. Existing MRP research has focused on identifying methods that produce population and sub-population estimates with high precision --- narrow prediction intervals for the population estimates at the multilevel regression stage, and low bias --- where the prediction for the population estimate is close to the assumed truth (e.g., \citeA{buttice2013, downes2018}). 

This is beneficial because it allows the readers to separate the difference between methods that improve bias and methods that improve precision. In the context of survey work, it might be particularly important to avoid biased estimates in some applications. However, the calculation of bias requires knowledge of the underlying truth and so can only be used in simulation studies and some special contexts (e.g., political science applications such as post-election results as in \citeA{isakov2021}). In this manuscript, our aim is to identify the usefulness of PSIS-LOO and WTD-PSIS-LOO, which do not require knowledge of the underlying truth and thus can be used to validate models in real-world settings. To facilitate easier comparison, we choose a single proper scoring rule that uses the underlying truth of the simulation to estimate the `goodness' of a predictive model. As we are comparing probabilistic estimates of a prediction, we choose the interval scoring rule in \citeA{gneiting2007}:
\begin{equation}
    S^{int}_\alpha (l, u; x) = (u - l) + \frac{2}{\alpha} (l - x) \mathbbm{1}\{ x < l \} + \frac{2}{\alpha} (x - u) \mathbbm{1}\{ x > u \}.
\end{equation}
where $\alpha$ is the threshold for the $(1 - \alpha) \times 100\%$ prediction interval $[u,l]$, $u$ and $l$ are the upper and lower bound of the prediction interval of the MRP estimates at both the population or the individual level and $x$ is the population and individual truths from our simulations. The interval score rewards narrow prediction intervals and penalizes when an observation misses the interval. Hence, a lower value of the interval score indicates a good model fit. 

Our argument proceeds as follows. In section \ref{sec:varsel}, we compare PSIS-LOO and WTD-PSIS-LOO in the context of population estimands predicted using different sets of variables. We demonstrate that PSIS-LOO and WTD-PSIS-LOO do not reflect the true model goodness at the population level. We continue our investigation by demonstrating that first, the best model for small area estimations (SAE) depends on the small area being estimated, and second, PSIS-LOO and WTD-PSIS-LOO are not sensitive to that either. In section \ref{sec:priorsel}, we compare two models with different regularizing priors in population and small area estimands, and again demonstrate a lack of consistency in PSIS-LOO and WTD-PSIS-LOO when used to score our models. Together, these suggest that LOO-based model ranking techniques may not always be suitable to evaluate the goodness of MRP estimates. We discuss the implication of this in section \ref{sec:diss}.

\section{Variable selection} \label{sec:varsel}
Our first challenge will consider the ability of PSIS-LOO and WTD-PSIS-LOO to identify models that are good at predicting population quantities (i.e., the population mean). We create a simulated scenario with two primary features. Firstly, a sampling design where the probability of inclusion is strongly related to two of our predictors, and two are weakly related. Second, we create the relationship between the outcome ($Y$) and categorical predictors ($\bm{X}$) where two are strongly predictive of our outcome and two are weakly related. Together this forms a two-by-two table for potential predictor variables shown in Table \ref{tab:reln} below. According to \citeA{little2003}, we would expect a model with $X_4$-omitted to produce biased MRP estimands and a model with $X_2$-omitted to produce less precise MRP estimates. Hence, for simplicity, we name the variables $X_4$ and $X_2$ as bias and precision variables accordingly. The remaining two are denoted \textit{ignorable} (not strongly related to the probability of inclusion or outcome) and \textit{inconsequential} (strongly related to sample inclusion but not the outcome). We name the variables accordingly for ease of reading. 

\begin{table}[htbp!]
    \centering
    \small
    \resizebox{0.95\textwidth}{!}{
    \begin{tabular}{ccccc}
    \multirow{2}{3.2cm}{Variable type}  &  \multirow{2}{3.5cm}{Weakly predictive of inclusion} & \multirow{2}{3.5cm}{Strongly predictive of inclusion}   \\ \\
    \midrule
    \multirow{2}{3.2cm}{Weakly predictive of the outcome} & \multirow{2}{3.5cm}{\centering $X_1$ \\ {\footnotesize(ignorable)}} & \multirow{2}{3.5cm}{\centering $X_3$ \\ {\footnotesize (inconsequential)}} \\ \\
    \multirow{2}{3.2cm}{Strongly predictive of the outcome} & \multirow{2}{3.5cm}{\centering $X_2$ \\ {\footnotesize(precision)} } & \multirow{2}{3.5cm}{\centering $X_4$ \\ {\footnotesize(bias)}} \\[1.8ex]
    \end{tabular} } 
    \caption{\em Relationship of variables with the simulated outcome and probability of inclusion}
    \label{tab:reln}
\end{table}

\subsection{Simulation design I}\label{sec:simdesignI}
We use 100 simulated iterations. For each, we create a new population of size $N = 10,000$. We define the population consisting of four variables in the set $K = \{X_1, X_2, X_3, X_4\}$ with $k$ denoting the $k$th member of the set. First, we generate four continuous variables from a normal distribution with a mean zero and a standard deviation of two. Then, the probability of outcome and inclusion probability were set to have an inverse-logit relationship with the sum of the continuous variables:
 \begin{align}
      \text{Probability of outcome:  Pr}(y_i = 1) &= \text{logit}^{-1}(0.1X_1 + 1X_2 + 0.1X_3 + 1X_4) \label{eq:yprob} \\
      \text{Inclusion probability: Pr}(I) &= \text{logit}^{-1}(0.1X_1 + 0.1X_2 + 1X_3 + 1X_4) \label{eq:inclprob}
 \end{align}
In our simulation, we denote 0.1 as the coefficient for a weak relationship and 1 for a strong relationship with the outcome and the inclusion probability. 

We create a binary outcome variable $Y$ through the binomial formulation using the Pr($Y$) obtained in equation (\ref{eq:yprob}). The inclusion probabilities Pr($I$) are generated as in equation (\ref{eq:inclprob}). Then, each $k$th continuous $X$ variable is discretized into $L^{(k)}= 5$ groups of equal range. The relationship between the outcome and inclusion probability were set up with the continuous variables instead of the discrete variables to allow stronger trends in their relationships. We use only the discrete variables and binary outcome in all of our models. \vspace{-0.2em}
 
\begin{figure}[H]
\centering
    \includegraphics[width=0.72\textwidth]{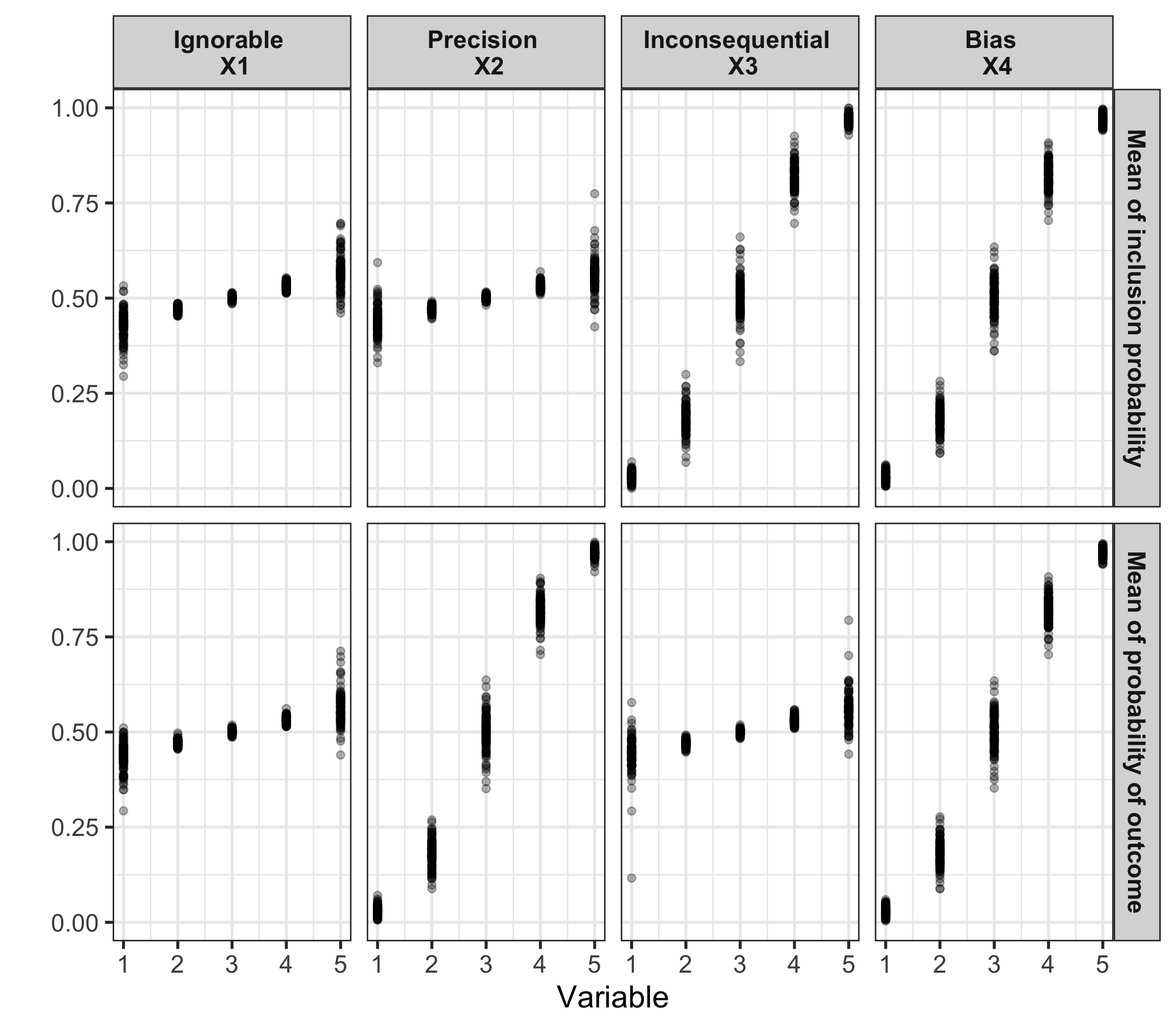}\vspace{-0.5em}
    \caption{\em Mean of inclusion probability and mean of probability of outcome for each variable in 100 iterations from simulation design I} \vspace{-0.5em}
    \label{fig:prob_plot} 
\end{figure} 
 
Now, we have generated the population with a binary outcome, discrete variables and continuous inclusion probability. Overall, our simulation setup has 100 different populations of size $N = 10,000$ and samples of size $n = 1,000$. To do this, we take a sample of size 980 according to the inclusion probability, and the remaining $L^{(k)} \times K = 20$ are sampled by enforcing each of the five levels for the four variables are included in the sample. We do this to avoid running into issues when constructing weights for a level that does not exist in the sample. We repeat the steps above for a 100 times to construct different populations and samples each time. This is considered the super-population approach, as we generate a different population each time. We employ this approach to test the efficacy of PSIS-LOO and WTD-PSIS-LOO for different simulated populations. 
 
For each sample, we then fit an MRP model consisting of two stages as in equation (\ref{eq:mlr1}) - (\ref{eq:mlr3}). The Bayesian multilevel logistic model is fitted through the probabilistic programming language \textit{Stan} \cite{carpenter2017, stan2019} as called from R \cite{cmdstanr, Rsoft}. A total of 15 different models were fitted by regressing the binary outcome on all combinations of the $\bm{X}$ variables ($2^4 - 1$, excluding the null model).\footnote{A summary of all the models is provided in the appendix.} We then poststratify the predicted population quantities from the models using the formula as in equation (\ref{eq:poststrat}). 

To assess the performance of PSIS-LOO and WTD-PSIS-LOO in different MRP settings, for each sample, we calculate PSIS-LOO values for each model using the \texttt{loo} package in R. The WTD-PSIS-LOO is calculated by using raked weights through the population margins for the set of variables $K = \{X_1, X_2, X_3, X_4\}$ using the \texttt{rake} function from the \texttt{survey} package in R.\footnote{A function \texttt{loo\_wtd} is written to calculate the WTD-PSIS-LOO by using raked weights and \texttt{svytotal} function from the \texttt{survey} package \cite{lumley2015}. The \texttt{loo\_wtd} function and all corresponding code from this manuscript are available on Github.}

\subsection{Results}
To investigate variable selection between different groups of models, we examine the PSIS-LOO-based model score values against the interval score, the selected `goodness' measure of a model. We study these on three instances, at the individual level, MRP population estimand level and the small area estimand levels. We describe and discuss the findings in the following sections.

\subsubsection{Individual level}\label{sec:varindv}
\begin{figure}[!htbp]
    \centering
    \includegraphics[width=\textwidth, height=.5\textheight]{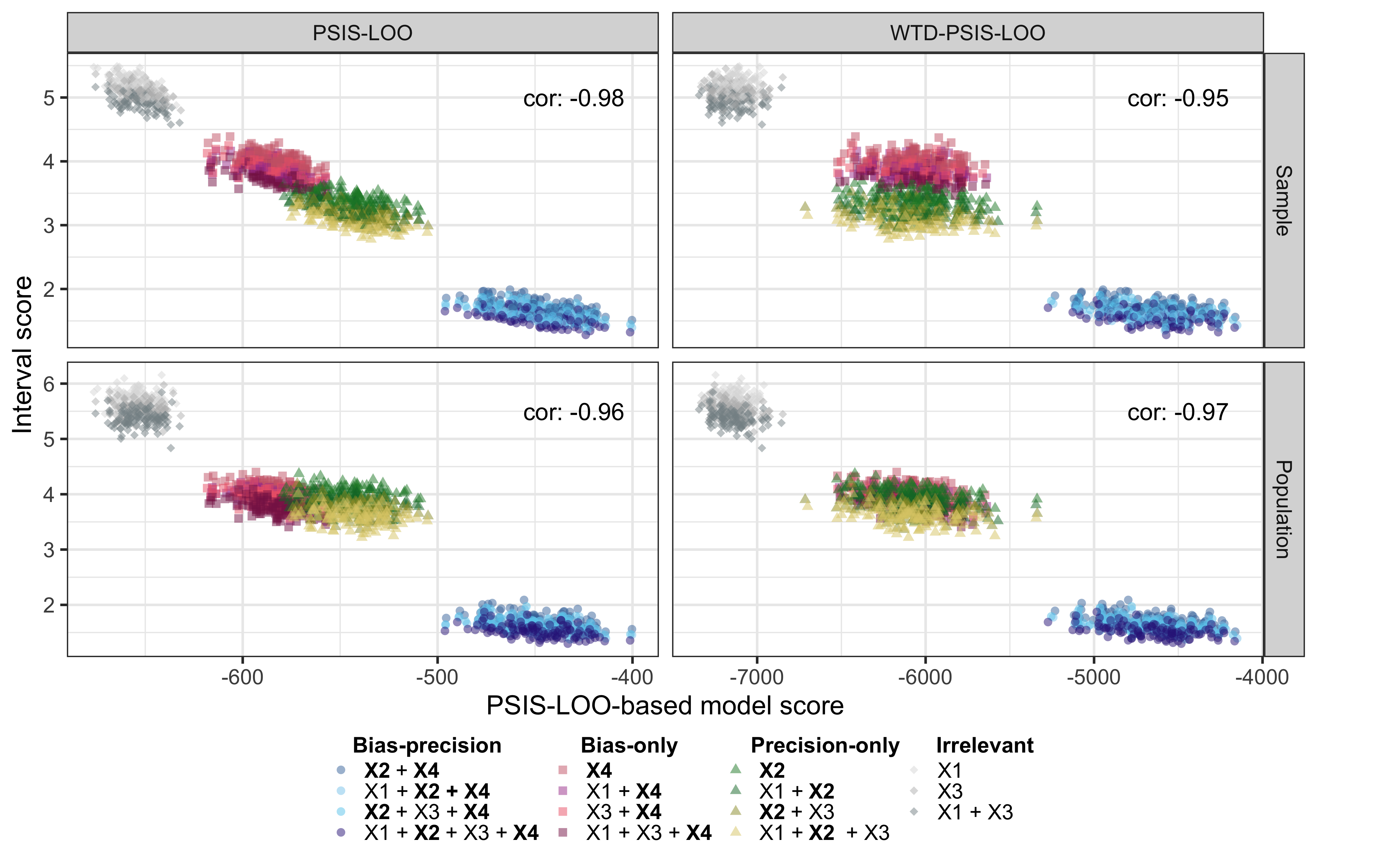}
    \caption{\em Interval score for average individual predictions ($y$-axis) and PSIS-LOO-based model scores ($x$-axis) in the sample and the population for 100 iterations. The interval score for the population individuals is calculated by multiplying individual raked weights by the mean interval score at the sample level. The four groups of colour represent different models (blue tone: models with both ``bias \&\ precision''; red tone: models with ``bias-only''; green tone: models with ``precision-only''; and grey tone: models with the ``irrelevant'' variables in them.) The $x$-axis on the left column panel represents PSIS-LOO as per \texttt{loo} package and the right column panel shows our proposed alternative WTD-PSIS-LOO. Ideally, the ranking of the models should be similar on both panels (which would be represented by a linear relationship between the two model scores).}
    \label{fig:elpd_indv}
\end{figure}

We first consider for each model, how well PSIS-LOO and WTD-PSIS-LOO accord with the indivi\-dual-level prediction accuracy in the sample (top row of Figure \ref{fig:elpd_indv}) as measured by the interval score (averaged over 100 iterations). If the PSIS-LOO approximately represents the true individual predicition goodness, we should see a negative and roughly linear relationship. As we see in the top left panel of Figure \ref{fig:elpd_indv}, this broadly is true. We see a clustering in model goodness, which will be a salient component of our results. The bias-precision models (blue-toned dots) are preferred by both scores, followed by the precision(-variable)-only models (green-toned dots), bias(-variable)-only models (red-toned dots) and the ``irrelevant'' variable models (grey-toned dots).

When we move to WTD-PSIS-LOO (top right column of Figure \ref{fig:elpd_indv}), the correlation between the two ranking criteria decreases slightly (from $-0.98$ to $-0.95$), and the bias- and precision-only variable models (red- and green-toned dots) are ranked similarly by WTD-PSIS-LOO, but not by the interval score. In other words, based on WTD-PSIS-LOO, both bias- and precision-only models are equally preferred. This might seem counterintuitive at first, but consider that we calculated the mean individual interval score in the sample -- not the population. The sampling design has changed the relative proportions of different $X_4$ (bias) levels in the sample (which led to the precision-only models being favoured over the bias-only models in the top left panel), but in the population $X_2$ and $X_4$ had similar distributions by creation. 

If we look at the mean individual interval score in the population (bottom row of Figure \ref{fig:elpd_indv}), we see that at the population level, the mean interval score for individuals are equally preferred for the bias- and precision-only models. As mentioned above, $X_2$ and $X_4$ in the population have similar distributions and this is reflected by the interval score measure, but not the unadjusted PSIS-LOO. When we adjust to WTD-PSIS-LOO using weights, both the interval score and WTD-PSIS-LOO agrees with the bias- and precision-only models, resulting in a huge overlap in the bottom right panel of Figure \ref{fig:elpd_indv} and a slight increase in correlation between the two ranking criteria (from -0.96 to -0.97).

\subsubsection{MRP estimand} \label{sec:varmrp}
\begin{figure}[!htb]
    \centering
    \includegraphics[width=\textwidth, height=0.4\textheight]{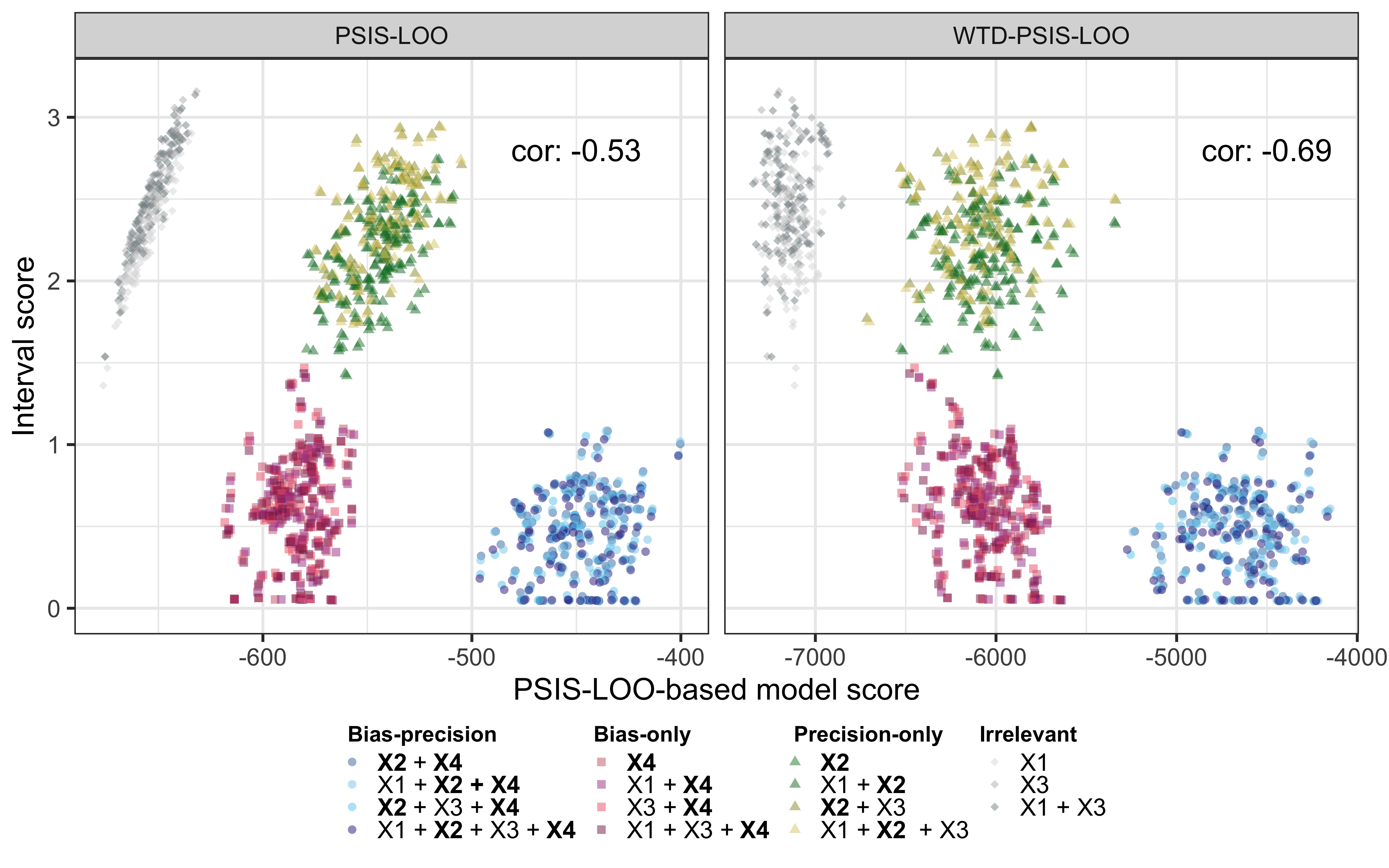}
    \caption{\em MRP population estimand interval score ($y$-axis) vs.\ PSIS-LOO-based model score ($x$-axis) in 100 iterations. The four groups of colour represent different models (blue tone: models with both ``bias-precision'' in it; red tone: models with ``bias-only''; green tone: models with ``precision-only''; and grey tone: models with the ``irrelevant'' variables in them.) The $x$-axis on the left panel represents PSIS-LOO as per \texttt{loo} package and the right panel shows the weighted alternative WTD-PSIS-LOO. Ideally, the ordering should be similar on both panels.}
    \label{fig:elpd_popn}
\end{figure}

Keeping in mind what the relationship should look like (top left of Figure \ref{fig:elpd_indv}), we now turn our attention to comparing PSIS-LOO and WTD-PSIS-LOO to the population interval score in Figure \ref{fig:elpd_popn}. On the $x$-axis, the full model (blue-toned dots) are preferred based on all three measures: PSIS-LOO, WTD-PSIS-LOO and interval score. Unfortunately, we see a large deviation between the ordering of the precision-only and bias-only models. While PSIS-LOO slightly prefers the precision-only model, the interval score reveals that the bias-only model is almost as good as the full model. We see a similar problem for the WTD-PSIS-LOO method, with a slight reordering of the bias- and precision-only models (correlation from -0.53 to -0.69), but not sufficiently to properly capture the true ordering. These bring suspicions that PSIS-LOO-based scores concerns only about the strength of relationship of the variable with the outcome, but not with the inclusion probability. We conduct extra simulations to confirm this and included them in the appendix. 

How can we interpret these findings? Firstly, there is good news. PSIS-LOO and WTD-PSIS-LOO do correctly suggest that the full model is the most appropriate for an MRP model in our simulation study. However, beyond this, the ordering is not correct. This is an issue for two reasons. One, there is no guarantee that we would always have the true model in the pool of models, and hence we are somewhat reliant on the ordering being fully correct. Secondly, there is a disadvantage to MRP models that is larger than they need to be, particularly if some of the variables are non-census variables and therefore require substantial work to be imputed in the population. In the case of our bias-only model, it is almost at good as the full model, but with one less variable. In some contexts, this model could be highly desirable. Our hypothesis of why this occurs lies with the aggregation stage of MRP, where errors can be averaged out like a beneficial Simpson's paradox. However, as our populations tend to the smaller samples (as in small area estimation), will PSIS-LOO get closer to identifying the true best model?

\subsubsection{Small area estimands}\label{sec:varsae}
MRP is a dual-purpose method. In addition to being interested in the population estimand, many MRP applications also aim to make small area estimates as well. Using the simulation setup I described in Section \ref{sec:simdesignI}, we examine the interval score for small area estimates in each level of the variable. As WTD-PSIS-LOO slightly corrects for our representation issue, we focus on WTD-PSIS-LOO in this section.

To examine small area estimation, we poststratify each model to each level of the variable instead of to the population. Let $K$ be the set of variables we wish to do small area estimations on, such that $K = \{ X_1, X_2, X_3, X_4\}$. Denote $k$th member of this set as $X^{(k)}$. Then, we aim to identify whether WTD-PSIS-LOO correctly preserves the ordering of models for each $k$th variable. To do this, we need to modify the calculations of WTD-PSIS-LOO so it only represents the prediction error for the level $\ell$ (e.g., 18--25 years old) for the $k$th variable (e.g., age group) in a way that is comparable across simulation iterations (remembering that by chance each sample will likely have a different sample size in each cell). 

We do this by taking the mean of the individual elpd values obtained through WTD-PSIS-LOO, but now only where $i \in S$ where $S$ is the set of individuals that are in our target small area level $\ell$ for each $k$th variable in $K$. Now, our $\theta_S$ in equation (\ref{eq:poststrat}) corresponds to the estimand for each level of the four variables. 

To examine each of the model validation scores, we evaluate WTD-PSIS-LOO and interval score at each of the variable levels. We take the mean of WTD-PSIS-LOO values at each small area instead of the sum due to the varying sample sizes in each group.\footnote{The mean sample size of each variable level for all 100 iterations are documented in the appendix.} Let $\xi_\ell$ be the set of individuals where $X_k = \ell$ and $\ell$ be the $\ell$th level for each covariate $X_k$, for $\ell = 1, \ldots, L^{(k)}$, and $L$ be the total number of levels where $L = \sum_k L^{(k)}$. The mean of WTD-PSIS-LOO values for each small area is then calculated by

\begin{align}
  \overline{\widehat{\text{elpd}}}_{\text{WTD-PSIS-LOO}[\ell]} = \sum_{i \in \xi_\ell}\frac{ \widehat{\text{elpd}}_{{\text{WTD-PSIS-LOO}[i]}}}{\norm{\xi_\ell}} \ \text{ for } i = 1, \ldots, n, \ \ell = 1, \dots, L^{(k)} \label{eq:saeCalc}
\end{align}

\noindent where $\norm{x}$ is the length of vector for all observations in $x$. We make similar small area calculations for the interval score by replacing $\text{elpd}_{\text{WTD-PSIS-LOO}}$ with interval score in equation (\ref{eq:saeCalc}).

With the calculated small area model scores, we plot the small area interval score and mean WTD-PSIS-LOO values for each variable when level $\ell = 5$ in Figure \ref{fig:elpd_sae_zoom}. We present the figure for all variable levels in the appendix. In Figure \ref{fig:elpd_sae_zoom}, we see that the best model for small area estimation differs based on the area being estimated. This suggests that MRP model validation might depend not only on the outcome being estimated, but also on the target estimand, and raises questions about how practitioners who aim to estimate multiple small areas, as well as an overall population should correctly identify the ``best'' model. In terms of which model is preferred, we see in the top right of Figure \ref{fig:elpd_sae_zoom} that the models without the precision variable (``bias-only'' and ``irrelevant'' models) have substantially worse interval scores for small area estimations of the precision variable (remember at the population this was not true). Conversely on the bottom right, for our bias variable, we see that models without the bias variable in them (``precision-only'' and ``irrelevant'') are substantially worse in terms of interval score. If we look at the $x$-axis only, WTD-PSIS-LOO does not seem prefer any of the models for our ignorable and inconsequential variables. When we look at the precision variable (top right plot), in terms of WTD-PSIS-LOO, although there are some preference for the models with the precision variable in them, the distinction across all models is less clear. This pattern is similar for the bias variable.

\begin{figure}[H]
    \centering
    \includegraphics[width=.95\textwidth]{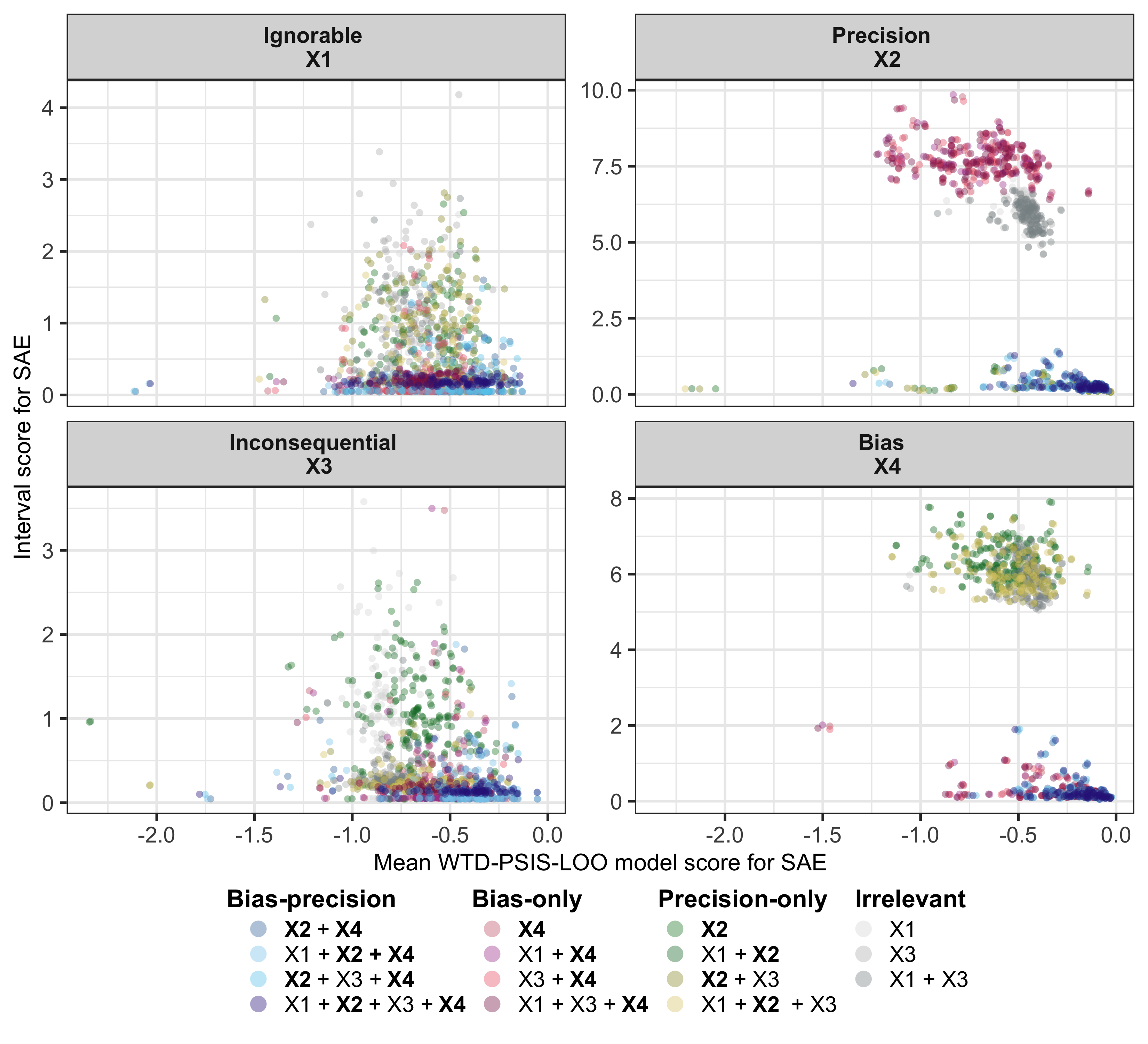}\vspace{-1.5em}
    \caption{\em Small area estimations of interval score ($y$-axis) vs.\ mean WTD-PSIS-LOO scores ($x$-axis) of the 5th level ($\ell = 5$) for each of the variables $X_1, X_2, X_3$ and $X_4$ for all 15 models in 100 iterations. The mean WTD-PSIS-LOO values of each level of the variables are as calculated through the \texttt{loo} package. The four groups of colour represent different models (blue tone: models with both ``precision\ \&\ bias'' in it; red tone: models with ``bias-only''; green tone: models with ``precision-only''; and grey tone: models with the  ``\textit{irrelevant}'' in them. The horizontal panels show each variable and the vertical panels represent the levels of the variables.}
    \label{fig:elpd_sae_zoom}
\end{figure}

We also see a similar problem as we saw in the population estimates when it comes to the usefulness of WTD-PSIS-LOO. While the best model differs for different variables, the actual ordering does not relate to the mean WTD-PSIS-LOO at all (for a particularly alarming example see the bottom-left figure). This suggests that a good MRP model for SAE differs based on the small area being estimated. WTD-PSIS-LOO, however, is unable to distinguish between the different models with distinct inclusion of the variables at the small area level, as shown in all of the panels in Figure \ref{fig:elpd_sae_zoom}. In the next section, we investigate if this challenge is specific to variable selection or if WTD-PSIS-LOO would be suitable when comparing models with the same variables but different priors. 

\section{Prior selection} \label{sec:priorsel}
To investigate the efficacy of PSIS-LOO-based scores on using different priors in our model, specifically in estimating small areas, we adapt simulation design I in Section \ref{sec:simdesignI}. The main difference from the previous simulation design is that we follow the setup closely of \citeA{gao2021} and apply an autoregressive (AR) prior (also named structured prior) on our bias variable $X_4$. We further describe the change in simulation design in the following section. 

\subsection{Simulation design II} \label{sec:simdesignII}

The idea of this simulation design is similar to the design in Section \ref{sec:simdesignI} but with some exceptions. Instead of generating our continuous variables from a normal distribution, we now simulate $\bm{X}$'s from a uniform distribution on $(-3, 3)$. We then discretize variables $X_1, X_2$ and $X_3$ into $L^{(k)} = 5$ groups, for $k = 1, 2,$ and $3$. For $X_4$, we discretize it into $L^{(k)} = 12$ groups, for $k = 4$ as in \citeA{gao2021} to allow the investigation of MRP prediction in smaller group sizes. We do this to test the efficacy of different priors and their effect on PSIS-LOO-based scores in SAE. As in Section \ref{sec:simdesignI}, we now have four variables, one strongly related to the probability of inclusion and outcome, the bias variable ($X_4$), one strongly related to the outcome, the precision variable ($X_2$) and two that are ignorable ($X_1$) and inconsequential ($X_3$) to our MRP predictions.

In estimating small area outcomes using MRP, we want to generate a stronger relationship between the variable and outcome, to investigate the effectiveness  of the use of prior. Therefore, instead of creating coefficients for $X_4$ as a random draw from a zero-mean normal distribution with a standard deviation of 2, we simulate a smooth relationship as the levels increase, mimicking the data simulation as in \citeA{gao2021} to investigate the improvements of the use of structured prior:

\begin{align}
      \text{Probability of outcome:  Pr}(y_i = 1) &= \text{logit}^{-1}\left(-\,1 + 0.05X_1 + 0.5X_2 + 0.05X_3 + 1.5f(X_4)\right) \label{eq:yprob2} \\
       f(x) &= 0.7 - 3\,\text{exp}\left(-\,\frac{x}{0.2}\right), x \in (-3, 3), \label{eq:fx3}
 \end{align}
where $f(x)$ in equation (\ref{eq:fx3}) is the increasing-shaped preference for the outcome as in \citeA{gao2021}.

\begin{figure}[H]
    \includegraphics[width=0.6\textwidth]{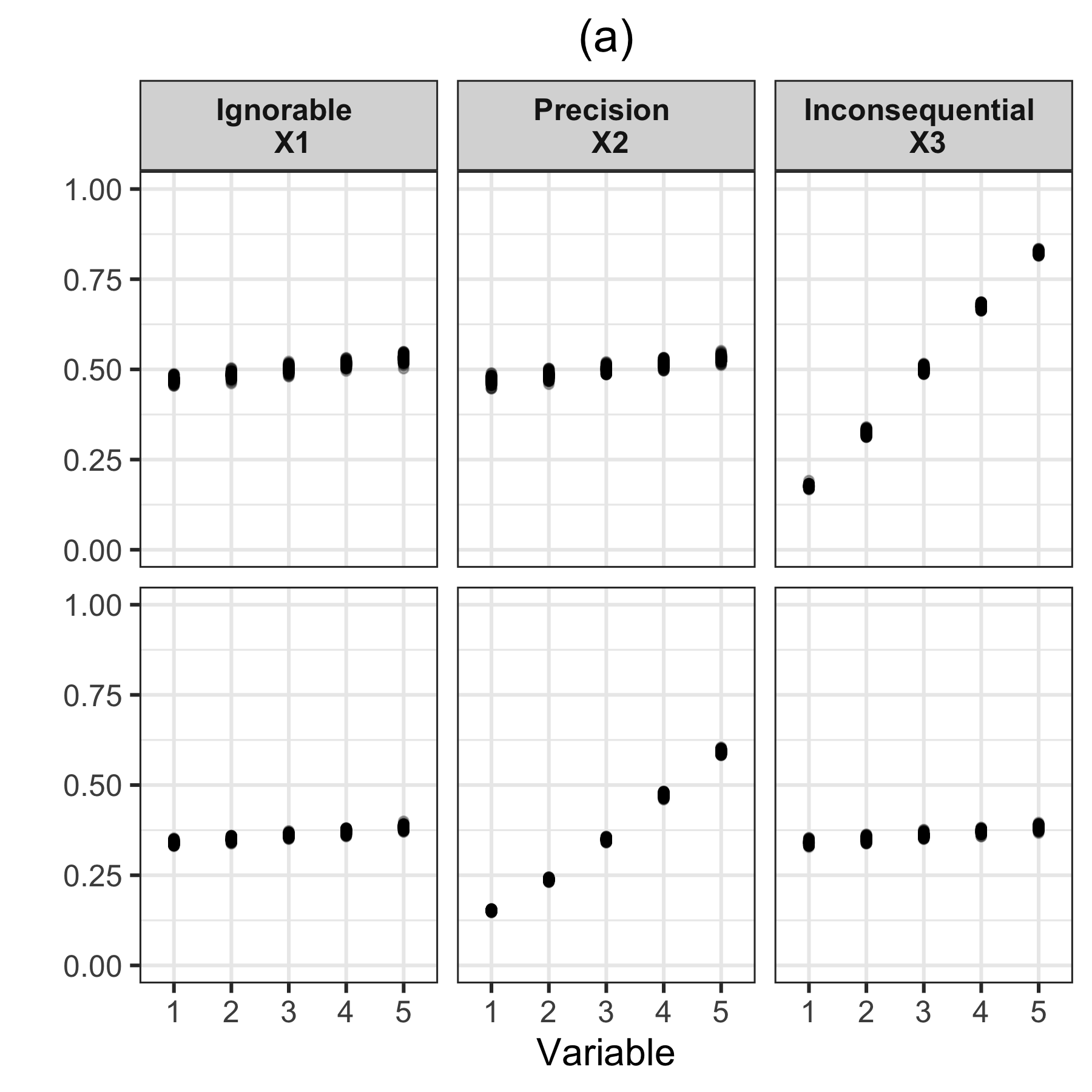} \includegraphics[width=0.33\textwidth]{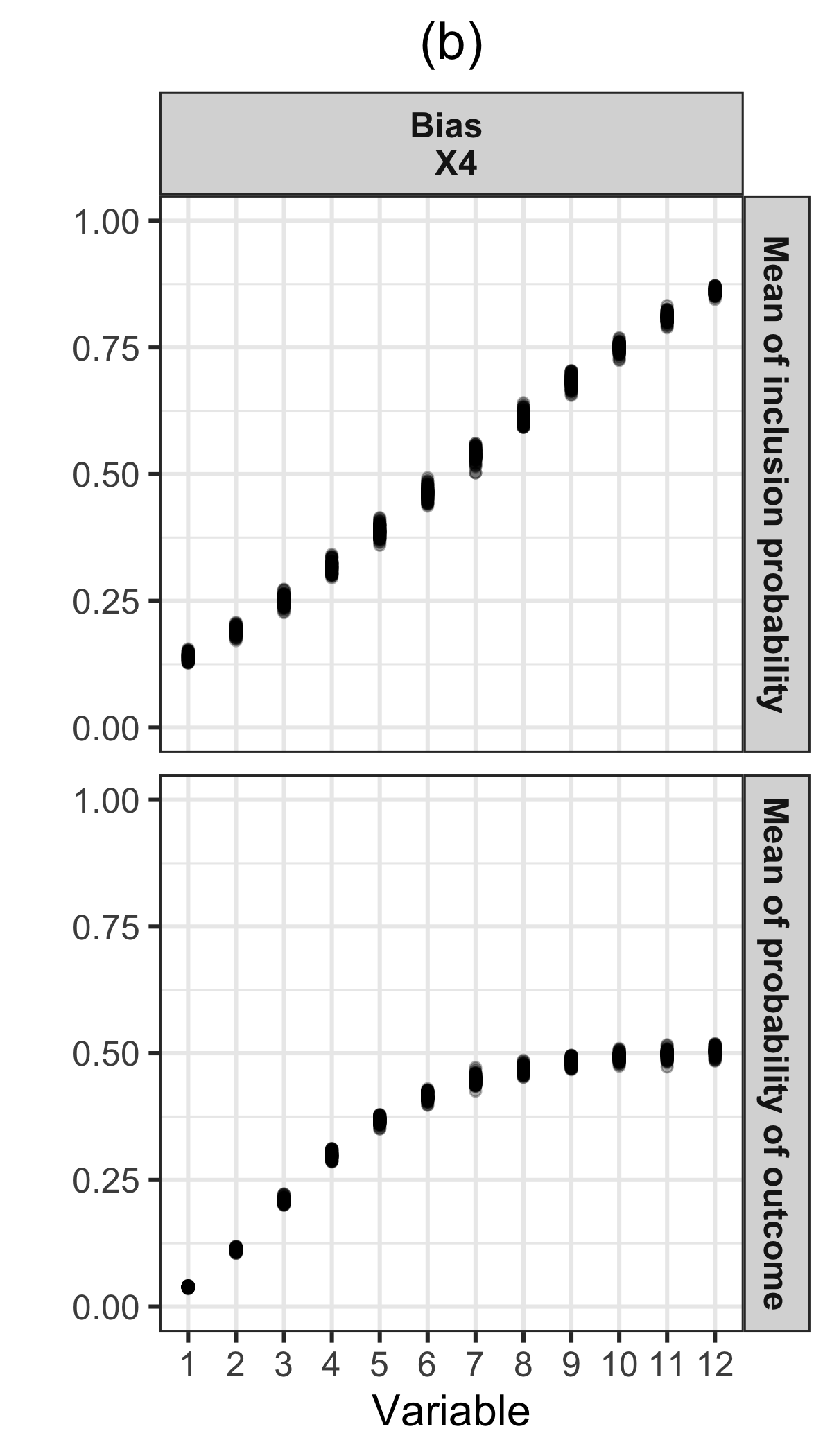}
    \caption{\em (a) Mean of inclusion probability and probability of outcome for all five levels of variables $X_1$, $X_2$ and $X_3$ in 100 iterations (b) Mean of inclusion probability and probability of outcome for all 12 levels of bias variable $X_4$ in 100 iterations}
    \label{fig:probSAE1}
\end{figure}

Then, we generate a binary outcome $Y$ using the new probability of outcome in equation (\ref{eq:yprob2}). Now, we have a binary outcome, three variables $X_1, X_2$ and $X_3$ with 5 levels and a bias variable $X_4$ with 12 levels. To motivate small-area estimations, we use a sample size of $n = 500$ to investigate the efficacy of PSIS-LOO-based scores on SAE of MRP predictions. As in the earlier simulation design, the sample is generated by taking the size of $(n - L)$ according to each of their inclusion probability. The remaining $L$ samples are simulated by making sure each of the variable levels is included in the sample. We demonstrate and discuss the results of sample size $n = 1000$ in the appendix. 
Using Stan, we fit the multilevel model similar to equation (\ref{eq:mlr1}) but for each small area $\ell$ for the $k$th variable in the set of variables $K$:
\vspace{-0.5em}
\begin{align}
    Pr(y_i = 1) &= \text{logit}^{-1}\left(\beta_0 +  \sum_{k \in K} \sum_{\ell^{(k)}}^{L^{(k)}} \alpha^{(k)}_{\ell[i]} \right) \quad \text{for } i = 1, \ldots, n; \ \ell^{(k)} = 1, \ldots, L^{(k)}.  \label{eq:yfitted2} 
\end{align}

Instead of the base normal prior distribution used in MRP as in simulation design I, we assume an AR prior distribution for our bias variable $X_4$:
 \vspace{-1.2em}
 \begin{align}
     \alpha^{(X_4)}_1 | \rho, \sigma^{(X_4)} \sim \mbox{normal}(0, \frac{1}{\sqrt{1 - \rho^2}} \sigma^{(X_4)}) \\
      \alpha^{(X_4)}_\ell | \alpha^{(X_4)}_{\ell - 1}, \ldots, \alpha^{(X_4)}_1, \rho, \sigma^{(X_4)} \sim \mbox{normal}(\rho\,\alpha^{(X_4)}_{\ell - 1}, \sigma^{(X_4)}).  \vspace{-0.5em}
 \end{align}

To consider the use of AR priors in SAEs, we compare the PSIS-LOO and WTD-PSIS-LOO preferred model from the previous section -- the ``full model'' that uses all four variables as the predictor, with and without the structured prior on the bias variable $X_4$ in the following section. 

\subsection{Results}
To investigate the use of AR priors, we compare the mean interval score and mean PSIS-LOO-based model scores for the full model with all variables, particularly with AR prior and with non-AR prior (base normal prior) on the bias variable. We first compute the difference in mean interval score and mean PSIS-LOO-based model scores (PSIS-LOO and WTD-PSIS-LOO) for all of the population individuals from simulation design II. We plot the difference in model scores for the population individual, MRP estimand and small-area estimand level. We calculate the difference in both scoring criteria to see if the model with AR prior performs better than the non-AR (base normal) prior, and if the interval score and PSIS-LOO agree on the differences. \vspace{-0.5em}

\subsubsection{Individual level}\label{sec:saepopn} \vspace{-1em}
\begin{figure}[!htb]
    \centering
    \includegraphics[width=\textwidth]{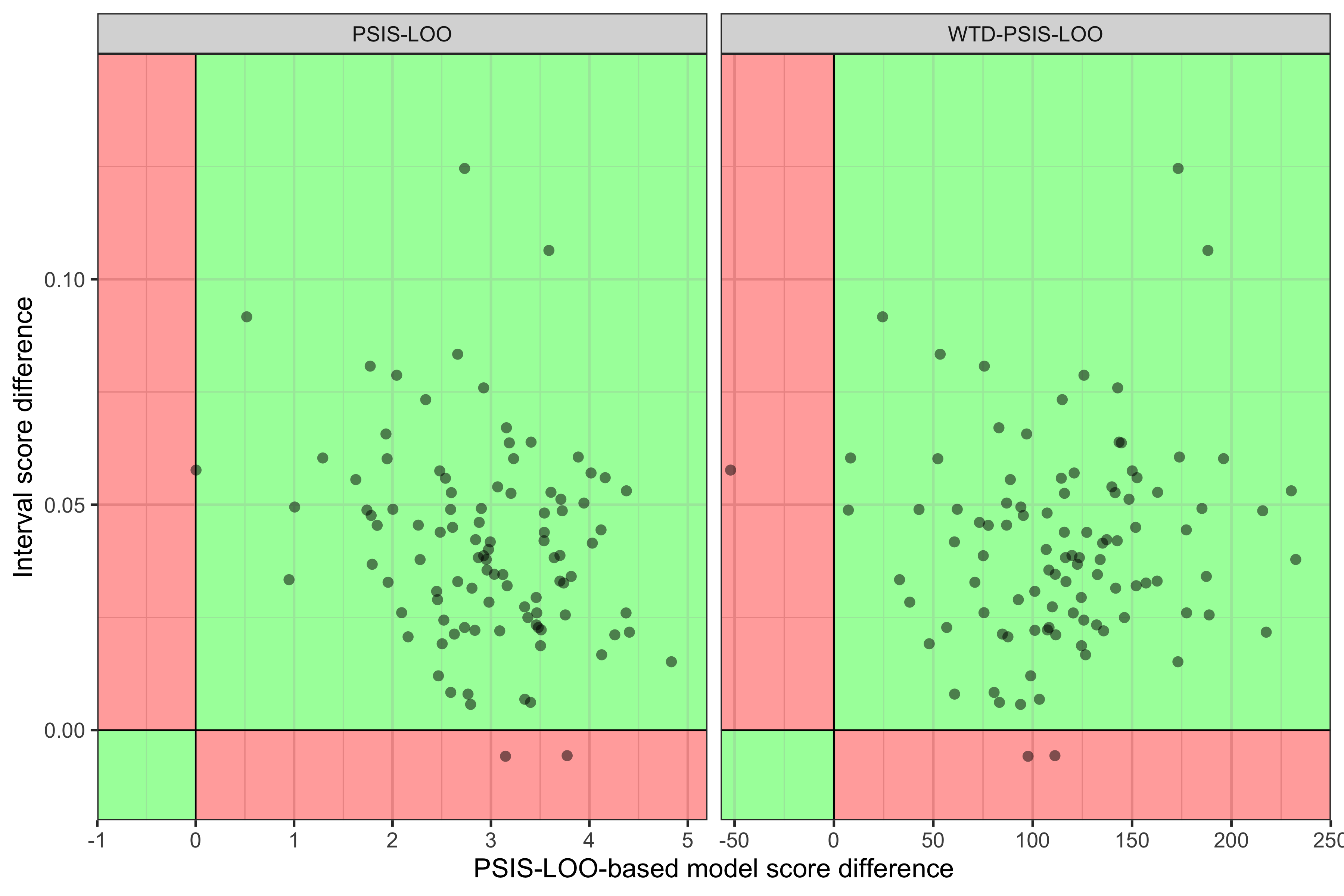} \vspace{-2em}
   \caption{\em The difference in mean interval score ($y$-axis) against the difference in mean PSIS-LOO-based scores ($x$-axis) for all population individuals in 100 iterations from simulation design II. This plot compares the difference between the full model that uses AR prior and non-AR prior on the bias variable, each using a sample size of 500. In each panel, the green quadrants (bottom left and top right) signify where interval score and PSIS-LOO-based model scores concur, and the red quadrants (top left and bottom right) signify where the two scores disagree. The figure shows that the majority of the points are scattered in the green quadrant for both PSIS-LOO and WTD-PSIS-LOO, where the two model scores agree.} \vspace{-1em}
\label{fig:AR_indv}
\end{figure}

We take the individual predictions at the population level and calculate the difference in the mean interval score and PSIS-LOO-based model scores in Figure \ref{fig:AR_indv}. The difference in mean interval score is calculated using $\text{diff}_{\text{IS}} = $ -(mean interval score (AR prior) $-$ mean interval score (non-AR prior)) where IS stands for interval score. We take the negative value of the difference as a smaller value of interval score indicates a better predictive model. We calculate the difference in mean PSIS-LOO-based models score using $\text{diff}_{\text{LOO-based}} = $ LOO score(AR prior) $-$ LOO score(non-AR prior) where LOO score indicates PSIS-LOO or WTD-PSIS-LOO values. 
 
The green quadrant signifies when PSIS-LOO corresponds to the perceived true goodness fit of the model where AR prior is working better. The red quadrants are where the model scores disagree in preference for the use of AR priors. At the individual level, we see that both scores perceive a similar preference for the AR prior model, where most points are scattered on the top right green quadrant. The similarity holds true for both PSIS-LOO and WTD-PSIS-LOO. \vspace{-0.5em}

\subsubsection{MRP estimand}\label{sec:saemrp} \vspace{-1em}
\begin{figure}[H]
    \centering
    \includegraphics[width=\textwidth]{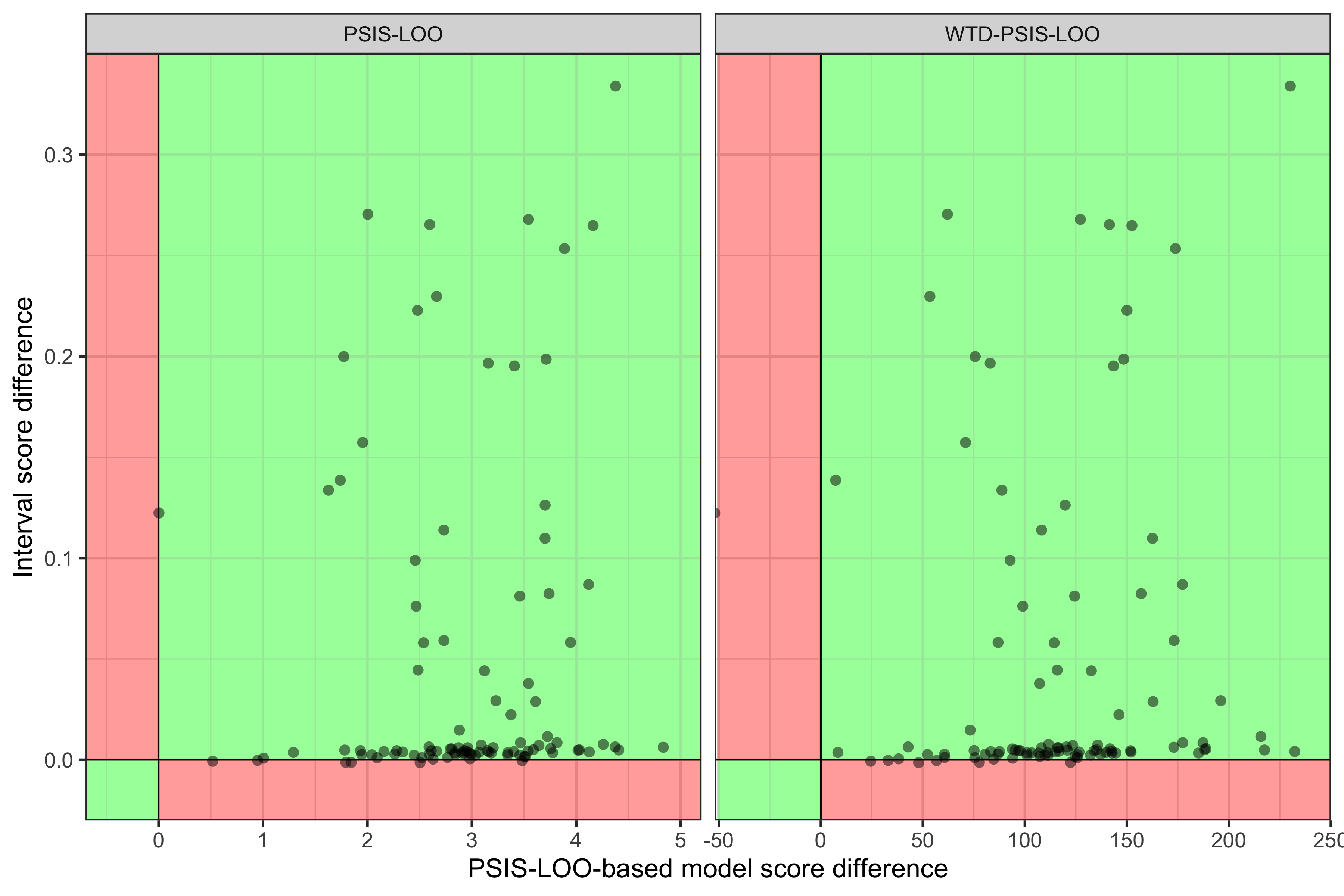} \vspace{-2em}
   \caption{\em The difference in interval score for MRP population estimand ($y$-axis) against the difference in mean PSIS-LOO-based scores ($x$-axis) in 100 iterations. This plot compares models using an AR prior and a non-AR prior on the bias variable, each using a sample size of 500. In each panel, the green quadrants (bottom left and top right) signify where interval score and PSIS-LOO-based model scores concur, and the red quadrants (top left and bottom right) signify where the two scores disagree. The figure shows that while the most of the points lie on the green quadrant, many of the points congregate around the horizontal intercept.}
\label{fig:AR_mrp}  \vspace{-0.7em}
\end{figure}

\hspace{\parindent}For Figure \ref{fig:AR_mrp}, we take the MRP population estimand predictions and calculate the difference in  interval score and difference in PSIS-LOO-based scores. The difference in mean for the models using AR prior and non-AR prior on the bias variable is calculated similarly as before. We see that at the MRP population estimand level, the AR prior model is only preferred by the true interval score in some of the cases, while for most iterations it signifies no difference in the use of AR prior. However, the PSIS-LOO-based model scores still suggest a difference between the models.

We then calculate the small-area estimates of interval score and PSIS-LOO-based model score difference in each of the levels of the bias variable $X_4$. As both WTD-PSIS-LOO and PSIS-LOO show similar results, we focus on WTD-PSIS-LOO for clarity. We plot the score difference in each small area of the bias variable $X_4$ between the two models with different priors in Figure \ref{fig:ARprior} below.

\subsubsection{Small-area estimands}\label{sec:saeprior} \vspace{-1em}
\begin{figure}[H]
    \centering
    \includegraphics[width=.9\textwidth]{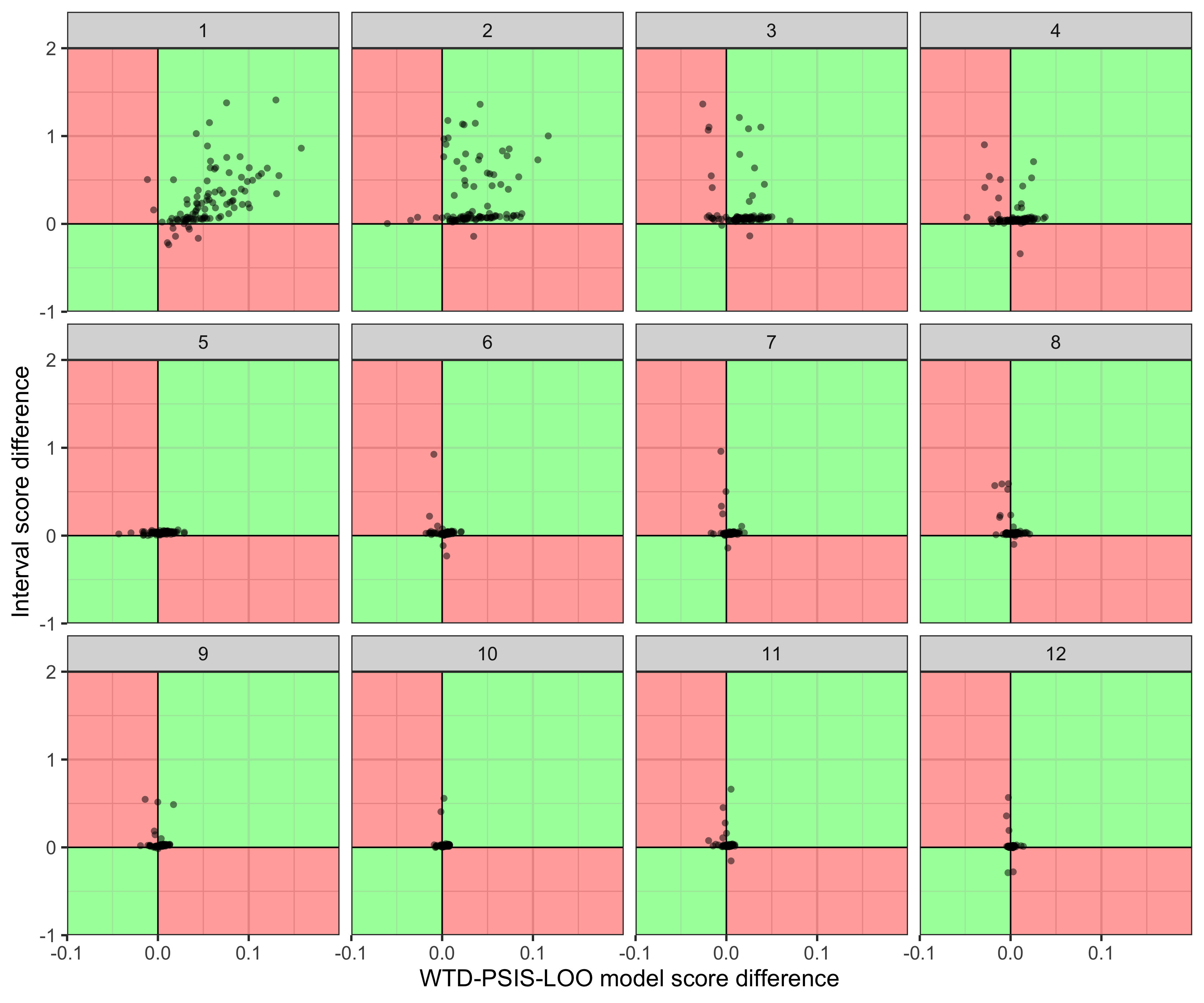}  \vspace{-0.5em}
    \caption{\em The small-area estimate of difference in interval score ($y$-axis) against difference in WTD-PSIS-LOO model scores ($x$-axis) for the bias variable $X_4$ in 100 iterations. This plot compares the full model that uses AR prior and non-AR prior on the bias variable, each using a sample size of 500. In each panel, the green quadrants (bottom left and top right) signify where interval score and WTD-PSIS-LOO values concur, and the red quadrants (top left and bottom right) signify where the two scores disagree. The two scores agree best when they are in a smaller group (in panels 1, 2, and 3), which is where structured prior in MRP is more effective than when compared to in bigger groups.} \vspace{-0.5em}
    \label{fig:ARprior}
\end{figure}

\hspace{\parindent}In Figure \ref{fig:ARprior}, we see that the small-area difference in PSIS-LOO values corresponds well to the interval score in levels 1, 2 and 3 of the bias variable $X_4$, as most of the points are in the green quadrant. This is where the group sample size is the smallest, averaging about 18 units in the sample.\footnote{The average group sizes of each variable level for 100 iterations are documented in the appendix.} The figure shows that as the variable level increases (so does the sample size due to the simulation setup), the points cluster around the origin. This signifies no difference in the use of AR prior on the higher variable level and increased sample sizes. This is coherent with the finding in \citeA{gao2021} where the structured prior works better for smaller group sizes than in bigger groups which has more stratification of levels in the variable. We find the structured prior is less useful when the total sample size increases to 1000 instead of 500, and display the results in the appendix.

\section{Discussion}\label{sec:diss}
Validating MRP is not straightforward. We focused here on using leave-one-out cross-validation and demonstrated three unique challenges.

The first challenge is {\em representation}:  MRP is typically used in settings where the sample is not representative of the broader population, but the leave-one-out cross-validation assumes that it is. In our simulation studies, this turned out to be the least impactful challenge. More alarming is the impact that aggregation has on the usefulness of PSIS-LOO-based measures. While in MRP, individual prediction error can average out in group averages, when using a cost function like expected log predictive density, these errors simply grow. We suspect this causes the inconsistencies in ordering that we observed in this work. Lastly, we observed that the best model depended on the target estimand (particularly population versus small area).

This all suggests that PSIS-LOO should be used with caution, if at all, to validate MRP models. The first challenge of representation, could be solved with a weighted alternative of PSIS-LOO. We had initially included an MRP version of this where weights are not available. However, it is eventually decided that the adjustment for representation was a relatively small contributor to the model validation issues. When using PSIS-LOO for MRP model validation, the weighted version helps slightly to rebalance the ordering, but as we found that is not the biggest challenge. 

The next major challenge is {\em aggregation}. As we saw in our results, using PSIS-LOO had the potential to suggest models that had more variables than needed for a reliable MRP estimate. In particular, PSIS-LOO was able to distinguish models that have \textit{both} variables that are strongly predictive of the outcome and inclusion. However, it is not able to differentiate between models that only have one of the variables. This has significant real-world implications for practitioners who are working with non-probability samples. If we rely only on PSIS-LOO, a model with non-census variables might get selected, which can be time-consuming if not impossible to create in the population. Our working hypothesis for this challenge is the default use of elpd as a cost metric in PSIS-LOO. Further work would be needed to confirm this, but if this is the case, it raises challenges for other elpd-based tools used with MRP, such as stacking \cite{ornstein2020}. 

The final challenge is {\em multiple goals}, which has been hinted at in previous work (e.g., \citeA{gao2021}), but is highlighted when we consider the best model for different small areas. The MRP model best suited to the job depends on the goal. That MRP is a multi-use tool, for both population and small area estimation, has always been a valuable feature. However, balancing multiple uses of this tool when it comes to validating and selecting the model has been challenging and requires further work. 

Overall, we have illustrated a cautionary tale. In our studies, PSIS-LOO and WTD-PSIS-LOO typically prefer a model that would give good MRP estimands. However, the ordering of all other models implies that our confidence in this depends on which models were included. We suggest potential reasons for this, but until these have been investigated further we suggest caution in interpreting PSIS-LOO-based model validation tools. Other strategies, such as comparisons to high-quality auxiliary surveys (where available) and known truths might still be our best tools when evaluating MRP-derived estimates. 

\bibliographystyle{apacite}
\bibliography{biblio}

\section{Appendix} \label{sec:appen}
\textbf{Abbreviations} 
AR - autoregressive;
elpd - expected log predictive density;
LOO - Leave-one-out cross-validation;
MRP - Multilevel regression and poststratification;
PSIS - Pareto smoothed importance sampling;
SAE - Small Area Estimation.

\subsection*{Types of models fitted}
In simulation design I, we expect $2^4 - 1 = 15$ different models for the combination of the four simulated variables. The 15 models are, in R programming syntax:

\begin{table}[!htb]
\begin{minipage}{.5\linewidth}
\begin{tabular}{rl}
 \multicolumn{2}{l}{\textbf{Bias-precision}} \\[3pt]
    1) & $Y \sim (1|X_2) + (1|X_4)$  \\[5pt]
    2) & $Y \sim  (1|X_1) + (1|X_2) + (1|X_4)$ \\[5pt]
    3) & $Y \sim (1|X_2) + (1|X_3) +  (1|X_4)$  \\[5pt]
    4) & $Y\sim (1|X_1) + (1|X_2) + (1|X_3) + (1|X_4)$  \\[5pt]
  \multicolumn{2}{l}{\textbf{Bias-only}} \\[3pt]
    5)& $Y \sim (1|X_4)$ \\[5pt]
    6)& $Y \sim (1|X_1) + (1|X_4)$ \\[5pt]
    7)& $Y \sim (1|X_3) + (1|X_4)$  \\[5pt]
    8)& $Y \sim (1|X_1) + (1|X_3) + (1|X_4)$ \\[5pt]
\end{tabular}
\end{minipage}\hfill

\begin{minipage}{.5\linewidth}
\begin{tabular}{ll}
 \multicolumn{2}{l}{\textbf{Precision-only}}  \\[3pt]
9) &$Y \sim (1|X_2)$ \\[5pt]
10) &$Y \sim (1|X_1) + (1|X_2)$ \\[5pt]
11) &$Y \sim (1|X_2) + (1|X_3)$ \\[5pt]
12) &$Y \sim (1|X_1) + (1|X_2) + (1|X_3)$ \\[5pt]
\multicolumn{2}{l}{\textbf{Irrelevant}} \\[3pt]
13) &$Y \sim (1|X_1)$ \\[5pt]
14) &$Y \sim (1|X_3)$ \\[5pt]
15) &$Y \sim (1|X_1) + (1|X_3)$
\end{tabular}
\end{minipage}
\end{table}

\subsection*{Change in strength of relationship of $X_2$ and $X_4$ in simulation design I}
\begin{figure}[H]
    \centering
    \includegraphics[width=.97\textwidth]{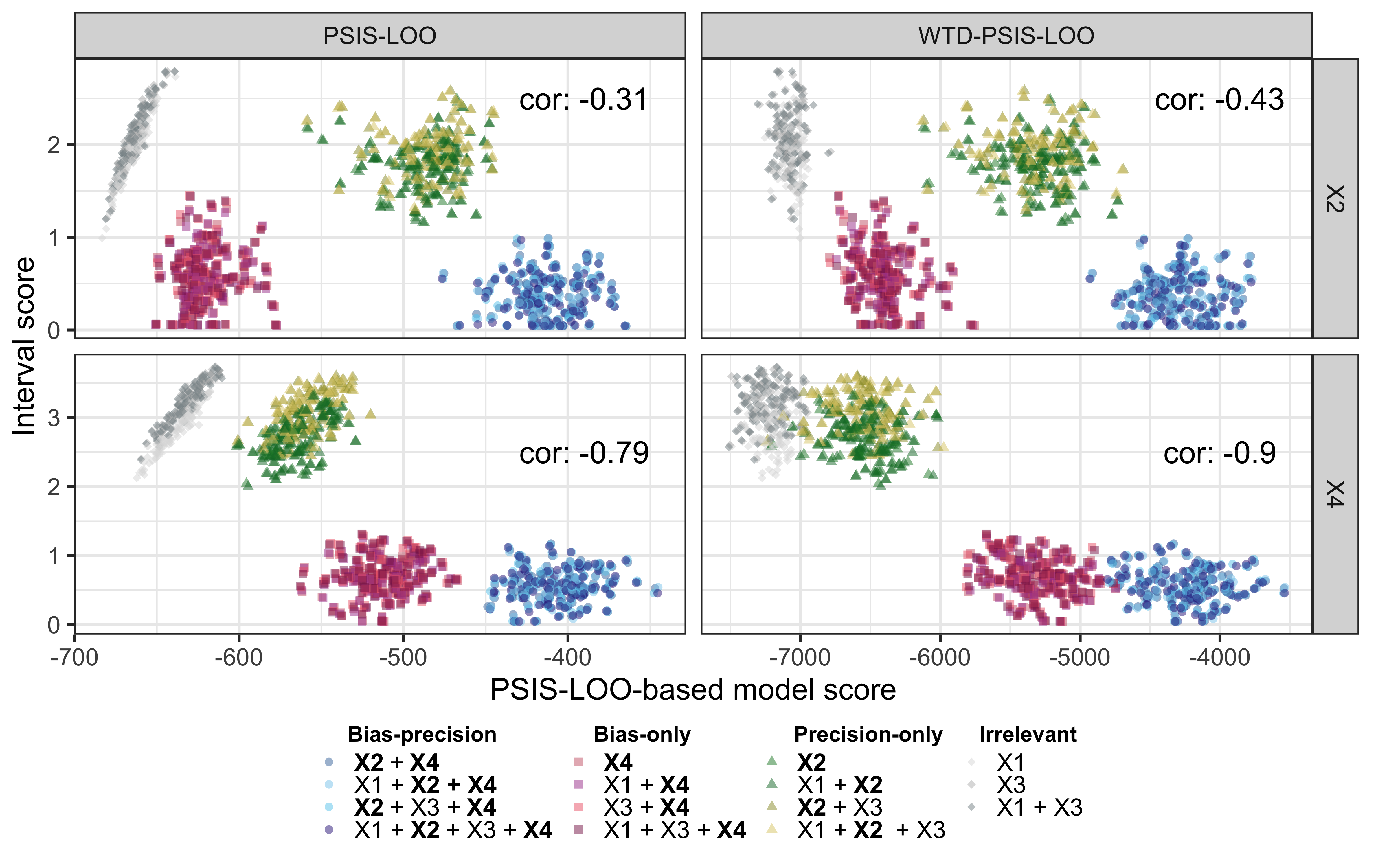}
    \caption{\em To examine if PSIS-LOO-based model scores mainly concerns about the relationship of a variable with the outcome but not the inclusion, we modify equation (\ref{eq:yprob}) in simulation design I. We change the strength of relationship between $X_2$ (precision) and $X_4$ (bias) with the outcome to 1.5 instead of 1, respectively, and keep the rest of the simulation design unchanged. We do this to evaluate if PSIS-LOO-based scores are sensitive to a stronger relationship between the variables and outcome. As a result, we see on the top panel that when the strength of relationship of $X_2$ and the outcome is stronger, PSIS-LOO model scores rank the (green-toned) precision-only models second, right after the (blue-toned) full model with both bias and precision variables in them. On the bottom panel, when we increase the strength of relationship of $X_4$ and the outcome, we see that the (red-toned) bias-only models are ranked second. The interval score on the $y$-axis consistently prefers the models with both bias-precision and bias-only models. This confirms our hypothesis of PSIS-LOO-based scores rank mainly based on the strength of relationship with the outcome but not the inclusion, which introduces specific challenges when validating MRP models, discussed at length in this paper.}
    \label{fig:elpd_popn_X2X4}
\end{figure}

\subsection*{Average sample sizes in simulation studies}
\begin{table}[H]
\centering
\begin{tabular}{cccccc}
  Level/variable & 1 & 2 & 3 & 4 & 5 \\ 
\midrule \\[-1.8ex] 
   $X_1$ & 13 & 211 & 549 & 214 & 13 \\ 
   $X_2$ & 12 & 202 & 546 & 226 & 15 \\ 
   $X_3$ & 2 & 89 & 544 & 340 & 25 \\ 
   $X_4$ & 2 & 86 & 548 & 339 & 24 
\end{tabular} 
 \caption{\em Average group sizes of our 100 simulated samples of size $n = 1000$ for each of the variables $X_1, X_2, X_3$ and $X_4$ in simulation design I} 
\end{table} 

\begin{table}[H] \centering 
  \label{} 
\begin{tabular}{@{\extracolsep{5pt}} ccccccccccccc} 
 \multirow{2}{1.1cm}{Level / variable}  & 1 & 2 & 3 & 4 & 5 & 6 & 7 & 8 & 9 & 10 & 11 & 12  \\  \\
\midrule \\[-1.8ex] 
$X_1$ & $95$ & $98$ & $99$ & $103$ & $105$ & \grey{NA} & \grey{NA} & \grey{NA} & \grey{NA} & \grey{NA} & \grey{NA} & \grey{NA} \\ 
$X_2$ & $95$ & $96$ & $101$ & $100$ & $107$ &  \grey{NA} & \grey{NA} & \grey{NA} & \grey{NA} & \grey{NA} & \grey{NA} & \grey{NA} \\ 
$X_3$ & $39$ & $68$ & $100$ & $132$ & $161$ &  \grey{NA} & \grey{NA} & \grey{NA} & \grey{NA} & \grey{NA} & \grey{NA} & \grey{NA} \\ 
$X_4$ & $14$ & $18$ & $22$ & $27$ & $33$ & $39$ & $44$ & $50$ & $55$ & $61$ & $66$ & $71$ 
\end{tabular} 
  \caption{\em Average group sizes of our 100 simulated samples of size $n = 500$ for each of the variables $X_1, X_2, X_3$ and $X_4$ in simulation design II } 
\end{table} 

\begin{table}[H]
\centering
\begin{tabular}{@{\extracolsep{5pt}} ccccccccccccc}
 \multirow{2}{1.1cm}{Level / variable}  & 1 & 2 & 3 & 4 & 5 & 6 & 7 & 8 & 9 & 10 & 11 & 12  \\  \\ 
\midrule \\[-1.8ex]
$X_1$ & 189 & 194 & 202 & 206 & 209 & \grey{NA} & \grey{NA} & \grey{NA} & \grey{NA} & \grey{NA} & \grey{NA} & \grey{NA}  \\ 
  $X_2$ & 188 & 193 & 202 & 204 & 212 & \grey{NA} & \grey{NA} & \grey{NA} & \grey{NA} & \grey{NA} & \grey{NA} & \grey{NA}  \\ 
  $X_3$ & 76 & 134 & 200 & 264 & 325 &\grey{NA} & \grey{NA} & \grey{NA} & \grey{NA} & \grey{NA} & \grey{NA} & \grey{NA} \\
  $X_4$ & 26 & 34 & 44 & 54 & 66 & 78 & 89 & 101 & 112 & 123 & 132 & 141
\end{tabular}
  \caption{\em Average group sizes of our 100 simulated samples of size $n = 1000$ for each of the variables $X_1, X_2, X_3$ and $X_4$ in simulation design II }
\end{table}

\subsection*{Small-area estimation in simulation design I}
\subsubsection*{Mean PSIS-LOO scores}
\begin{figure}[H]
    \centering
    \includegraphics[width=.95\textwidth]{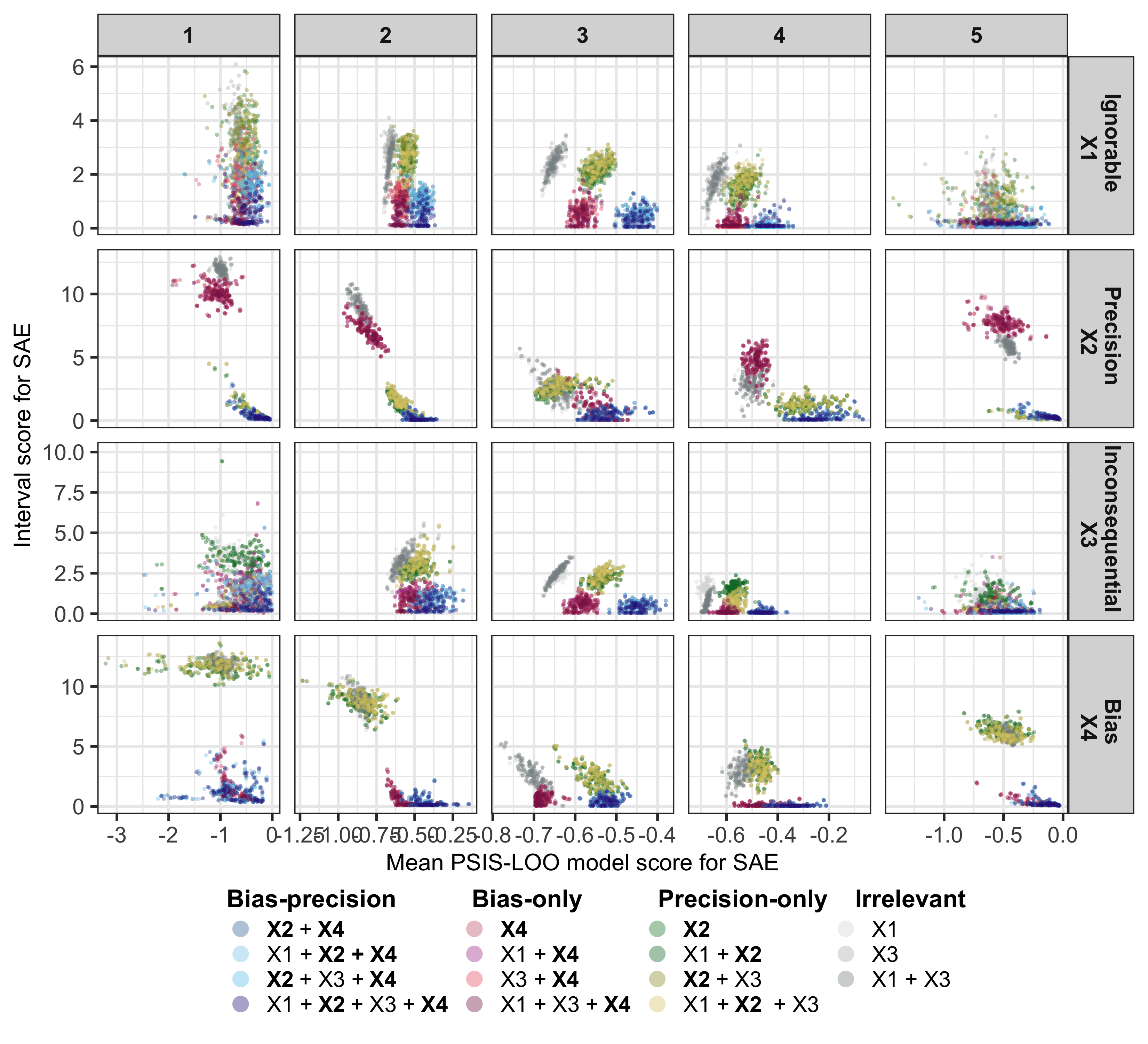}\vspace{-1.5em}
    \caption{\em Small-area estimates of interval score ($y$-axis) vs.\ mean PSIS-LOO scores ($x$-axis) of each level of the variables $X_1, X_2, X_3$ and $X_4$ for all 15 models in 100 iterations. The mean PSIS-LOO values of each level of the variables are as calculated through the \texttt{loo} package. The four groups of colour represent different models (blue tone: models with both ``precision\ \&\ bias'' in it; red tone: models with ``bias-only''; green tone: models with ``precision-only''; and grey tone: models with the  ``\textit{irrelevant}'' in them. The horizontal panels show each variable and the vertical panels represent the levels of the variables. We see that the preferred model for small area estimation differs based on the area being estimated. }
    \label{fig:elpd_sae_1000}
\end{figure}

\subsubsection*{Mean WTD-PSIS-LOO scores}
\begin{figure}[H]
    \centering
    \includegraphics[width=.95\textwidth]{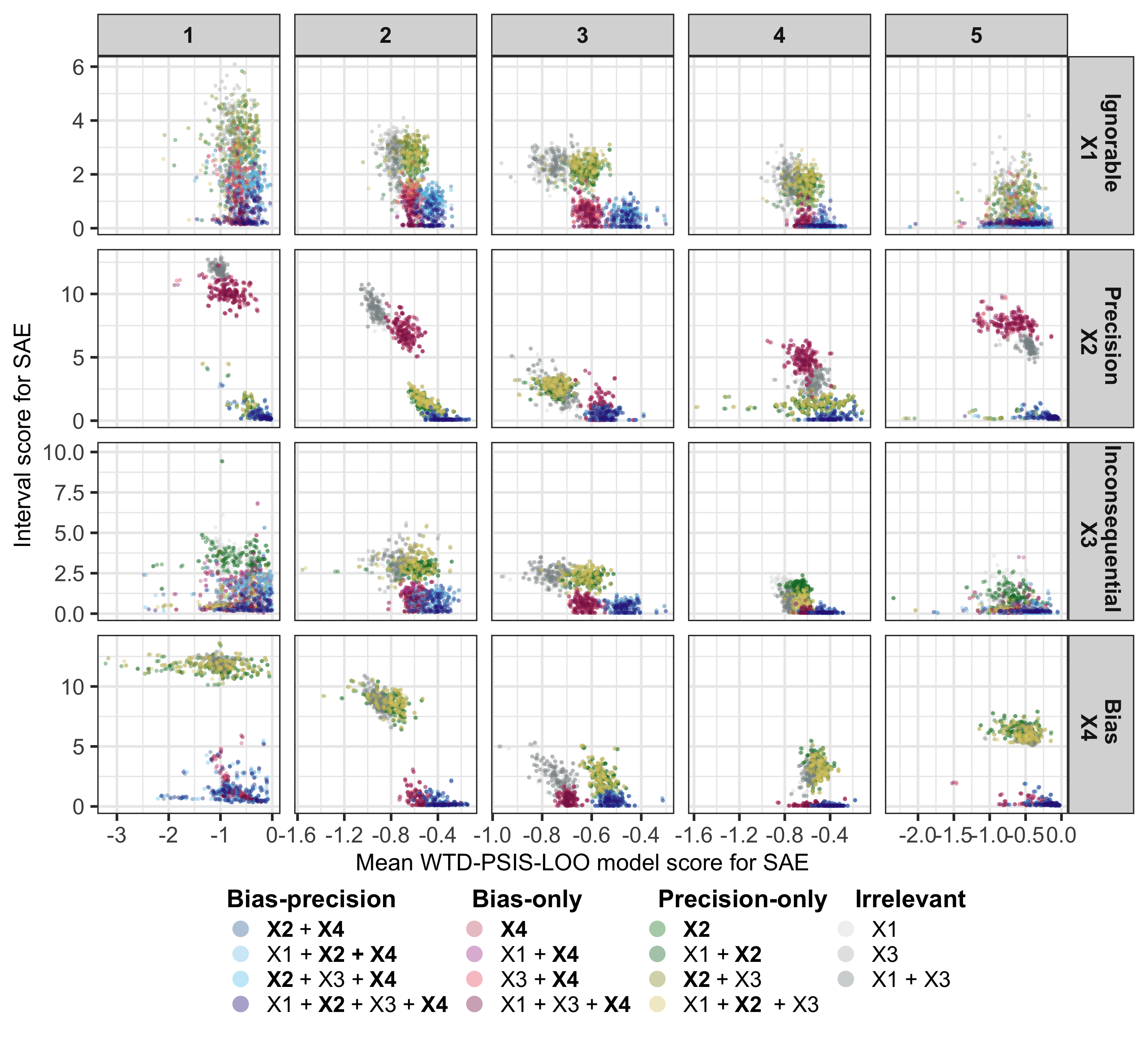}\vspace{-1.5em}
    \caption{\em Small-area estimates of interval score ($y$-axis) vs.\ mean WTD-PSIS-LOO scores ($x$-axis) of each level of the variables $X_1, X_2, X_3$ and $X_4$ for all 15 models in 100 iterations. The mean WTD-PSIS-LOO values of each level of the variables are as calculated through the \texttt{loo} package. The four groups of colour represent different models (blue tone: models with both ``precision\ \&\ bias'' in it; red tone: models with ``bias-only''; green tone: models with ``precision-only''; and grey tone: models with the  ``\textit{irrelevant}'' in them. The horizontal panels show each variable and the vertical panels represent the levels of the variables. We see a similar pattern to the PSIS-LOO scores where the best model for small area estimation vary based on the area being estimated.}
    \label{fig:elpd_sae_1000_wtd}
\end{figure}

\subsection*{Interval score, bias, and precision of MRP estimates in simulation design I}

\begin{figure}[H]
    \centering
     \includegraphics[width=.98\textwidth]{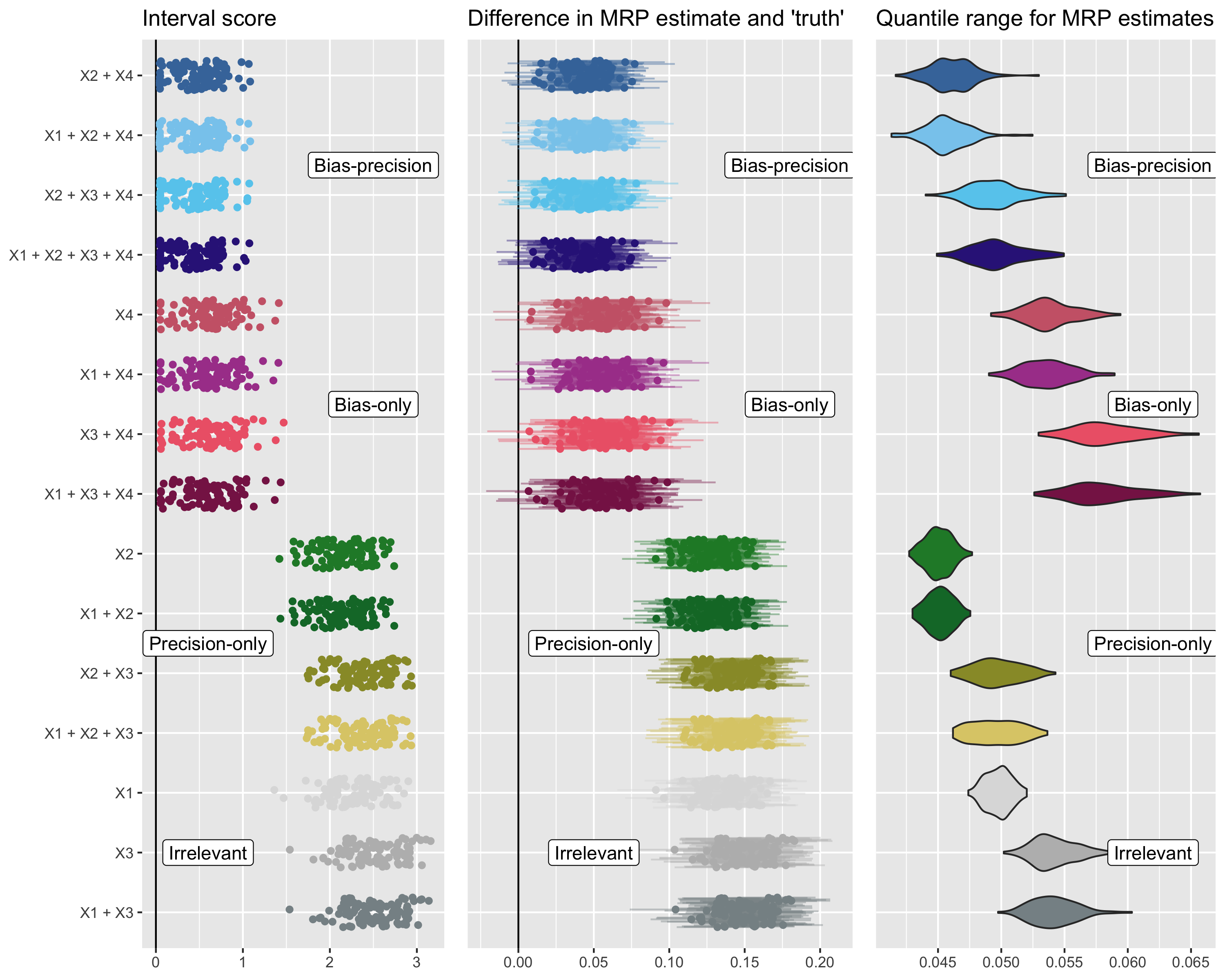}
    \caption{Comparison of the interval score, bias, and precision of the MRP estimates for the different models. The left panel shows the interval scores for the MRP estimates. The middle panel shows the difference between MRP prediction and the population truth (bias of MRP estimates). The points are the difference between the mean and ground truth and error bars are the difference between the 90\% prediction interval and the truth. The right panel shows the violin plot for the width of the 90\% prediction interval. Each point is calculated by taking the difference between 95\% and 5\% prediction interval value.}
    \label{fig:biasprec}
\end{figure}

\subsection*{Small-area estimates in simulation design II}
\begin{figure}[H]
    \centering
\includegraphics[width=\textwidth]{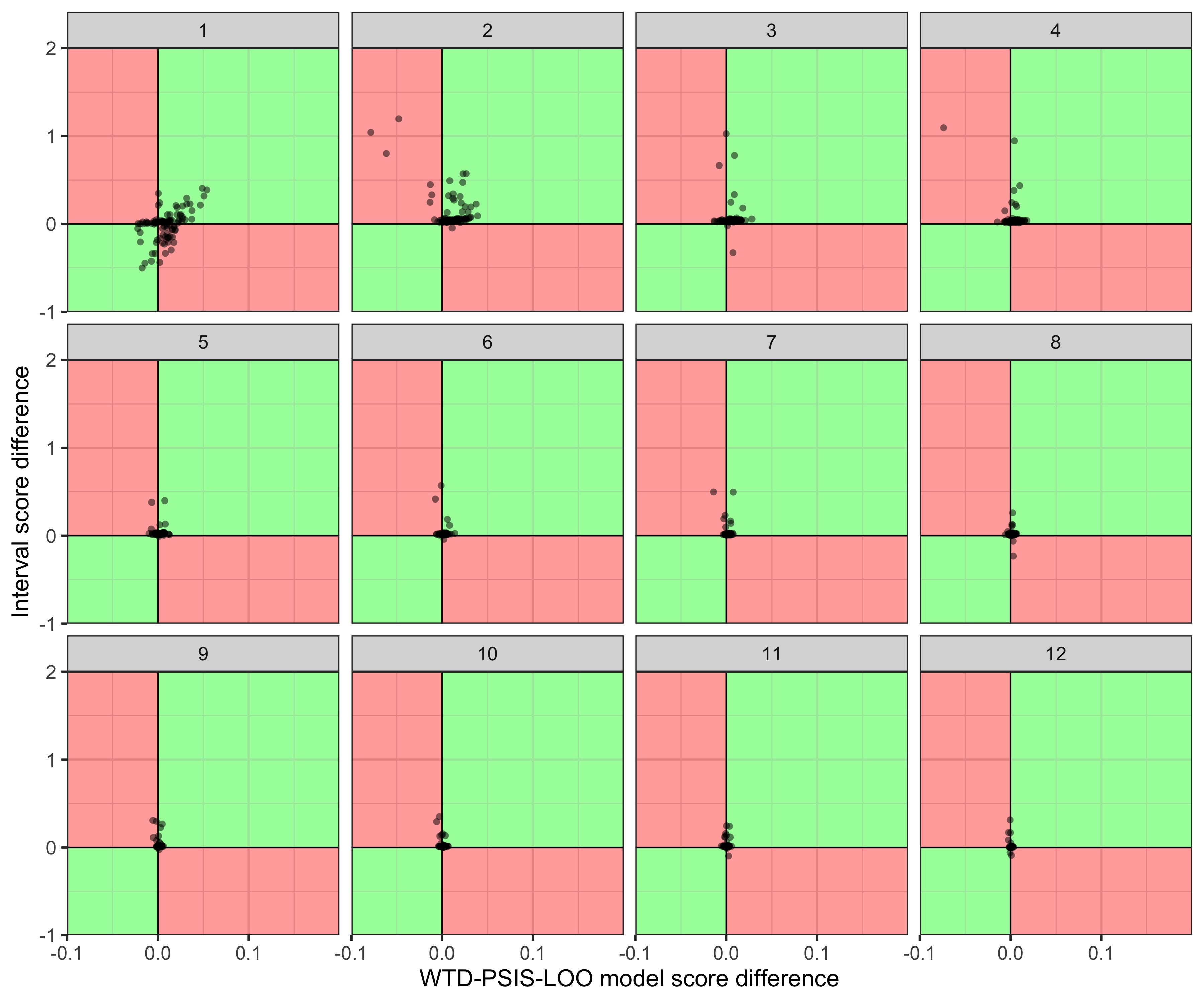}
    \caption{\em Similar to Figure \ref{fig:ARprior} but using a size of 1000 for each sample. The results appear to be less conclusive, where the use of AR prior does not make a huge difference, which is to be expected in larger sample size, as MRP is found to be the most useful in SAEs.}
    \label{fig:my_label}
\end{figure}
\end{document}